\documentclass[aps, prc, twocolumn, floatfix, nofootinbib, superscriptaddress, longbibliography, amsmath, amssymb]{revtex4-1}

\usepackage{mathrsfs}
\usepackage{graphicx}
\usepackage{epsfig}
\usepackage{bm}
\usepackage{bbold}
\usepackage{color}
\usepackage{float}
\usepackage{dcolumn}
\usepackage{multirow} 
\usepackage{hyperref}
\usepackage{color}
\usepackage{braket}
\usepackage{dcolumn}
\usepackage{multirow}
\usepackage[titletoc]{appendix}
\usepackage{cleveref}
\usepackage{url}
\usepackage{enumitem}
\usepackage{array}

\newcolumntype{d}[1]{D{.}{.}{#1}}

\begin{document}

\renewcommand{\arraystretch}{1.25}

\title{Neutrinoless double-$\beta$ decay from an effective field theory for heavy nuclei}

\author{C. Brase}
\affiliation{Technische Universit\"at Darmstadt, Department of Physics, 
64289 Darmstadt, Germany}
\affiliation{Max-Planck-Institut f\"ur Kernphysik, Saupfercheckweg 1, 69117 Heidelberg, Germany}

\author{J. Men\'endez}
\affiliation{Departament de Física Quàntica i Astrofísica, Universitat de Barcelona, 08028 Barcelona, Spain}
\affiliation{Institut de Ciències del Cosmos, Universitat de Barcelona, 08028 Barcelona, Spain}

\author{E. A. Coello P\'erez} 
\affiliation{Lawrence Livermore National Laboratory, Livermore, California 94550, USA}

\author{A. Schwenk} 
\affiliation{Technische Universit\"at Darmstadt, Department of Physics, 
64289 Darmstadt, Germany}
\affiliation{Max-Planck-Institut f\"ur Kernphysik, Saupfercheckweg 1, 69117 Heidelberg, Germany}
\affiliation{ExtreMe Matter Institute EMMI, Helmholtzzentrum f\"ur
Schwerionenforschung GmbH, 64291 Darmstadt, Germany}

\begin{abstract}
We study neutrinoless double-$\beta$ decay in an effective field theory (EFT) for heavy nuclei, which are treated as a spherical core coupled to additional neutrons and/or protons. Since the low-energy constants of the EFT cannot be fitted to data for this unobserved decay, we follow an alternative strategy to constrain these through a correlation with double Gamow-Teller transitions. This correlation was recently found to hold for shell-model calculations, energy-density functionals, and other nuclear structure models. We therefore first calculate the nuclear matrix elements for double Gamow-Teller transitions in the EFT for heavy nuclei. The combination of the EFT uncertainty with the correlation uncertainty enables predictions of nuclear matrix elements for neutrinoless double-$\beta$ decay for a broad range of isotopes with quantified uncertainties. Generally the EFT predicts smaller nuclear matrix elements compared to other approaches, but our EFT results are consistent with recent \textit{ab initio} calculations.
\end{abstract}

\maketitle

\section{Introduction}

Neutrinoless double-$\beta$ ($0\nu\beta\beta$) decay is one of the most promising processes to discover lepton number violation and thus physics beyond the Standard Model (BSM)~\cite{Avignone:2007fu,Dolinski:2019nrj}.
In this possible decay mode of an atomic nucleus, two neutrons decay into two protons while two electrons are emitted, effectively creating two leptons.
$0\nu\beta\beta$ decay is possible because neutrinos are neutral massive particles, therefore candidates for being their own antiparticles (Majorana particles).
In general, the decay can be triggered by BSM extensions that do not conserve lepton number, leading to a $0\nu\beta\beta$ rate proportional to some BSM parameter encoding such violation.
On the other hand, the $0\nu\beta\beta$ rate is also proportional to the nuclear matrix element (NME) that captures the many-body aspects of the nuclear decay.
Since the BSM mechanism responsible for $0\nu\beta\beta$ decay is unknown, reliable NMEs are needed to obtain information about BSM physics once $0\nu\beta\beta$ is detected. In the meantime, NMEs are used to constrain BSM scenarios based on the most stringent experimental limits~\cite{Ga16zero2beta,EXO:2019wmi,CUORE:2019jhp,GERDA:2020xhi}.
In the standard scenario where the decay is driven by the known light neutrinos, NMEs are also key to explore the physics reach of next-generation $0\nu\beta\beta$ experiments~\cite{LEGEND:2017syy,nEXO:2017hjq,CUPID:2019loe,DARWIN:2020adk,CROSS:2019dty,NEXT:2020cye} in terms of a combination of neutrino masses and mixing parameters.

However, the $0\nu\beta\beta$ NMEs relevant for $0\nu\beta\beta$ searches are not well known~\cite{En17NuclMEstatusandfuture}.
While NMEs can be predicted with good precision in lighter or moderately correlated nuclei~\cite{Pastore:2017ofx,Wang:2019hjy,Basili:2019gvn,Yao:2019rck,Belley:2020ejd,Novario:2020dmr,Yao:2020olm}, $\beta\beta$ emitters are in general medium-mass or heavy systems with complex nuclear structure.
Because of this, state-of-the-art calculations of $0\nu\beta\beta$ NMEs disagree up to a factor 3 and, in addition, all calculations may be subject to additional uncertainties~\cite{En17NuclMEstatusandfuture}.
One of the main limitations of current predictions of $0\nu\beta\beta$ NMEs is the difficulty of addressing theoretical uncertainties in a reliable and systematic manner.
Except in lighter systems~\cite{Pastore:2017ofx,Wang:2019hjy,Basili:2019gvn,Yao:2020olm} and very recent calculations using \textit{ab initio} methods~\cite{Yao:2019rck,Belley:2020ejd,Novario:2020dmr}, most NMEs have been obtained using nuclear models that rely to some extent on phenomenological adjustments~\cite{En17NuclMEstatusandfuture}. In turn these prevent the estimation of associated theoretical errors.
At present, the best NME uncertainty estimation within a model is to compare the different results obtained when varying the parameters and truncations of that model.

Effective field theories (EFTs) provide an alternative to calculate nuclear properties~\cite{Hammer:2019poc}.
By capturing the relevant degrees of freedom and encoding the symmetries of the problem, EFTs provide a systematic order-by-order expansion in a small parameter.
While in practice all EFT calculations truncate this expansion at a given order, the systematic character of the EFT allows one to quantify the theoretical uncertainty of the results. 
Previous work has shown that an EFT for heavy nuclei~\cite{Papenbrock:2010yg,Papenbrock:2013cra,Papenbrock:2015mdi,Chen:2017inn,Chen:2018jtd,Papenbrock:2020zhh} can describe well different properties, including electromagnetic~\cite{Co15ETrigrot,Co15ETvibuncqant,Co16nETvoo} and weak~\cite{Coe18Gamow-Teller} transitions.
This work is based on the same EFT used to explore beta ($\beta$) and two-neutrino $\beta\beta$ decays in Ref.~\cite{Coe18Gamow-Teller}, which considered a broad range of nuclei and showed that the measured decays agree well with the EFT predictions within uncertainties. In particular for two-neutrino $\beta\beta$ decay, six of the seven measured decays fall within the EFT uncertainties (see Fig.~5 in Ref.~\cite{Coe18Gamow-Teller}) and also the Gamow-Teller NMEs generally agree with the EFT predictions. Furthermore, the EFT successfully predicted the double electron capture half-life of $^{124}$Xe ~\cite{Perez:2018cly} before its recent discovery ~\cite{XENON:2019dti}. We take these results as a confirmation of the validity of the EFT for
the nuclei studied in this work.

In this paper, we extend the EFT for heavy nuclei to predict $0\nu\beta\beta$ NMEs.
In order to evaluate NMEs for any process in an EFT, the couplings associated with the relevant operator, usually called low-energy constants (LECs), need to be fitted to data.
However, since $0\nu\beta\beta$ decay has not been observed, such strategy cannot be applied in this case.
An alternative would be to fit the LECs to a theoretical matching calculation, which is also complicated in this case due to the heavy nuclei involved.
Instead, here we use the correlation recently found between double Gamow-Teller (DGT) transitions and $0\nu\beta\beta$ decay~\cite{Shimizu:2017qcy}.
This correlation is supported by approaches as diverse as the nuclear shell model (NSM)~\cite{Mene09nbb,Menendez:2017fdf}, energy density functional (EDF) calculations~\cite{Rodriguez:2012rv}, and the interacting boson model (IBM)~\cite{Barea:2015kwa}.
Interestingly, these models disagree in the prediction of specific NMEs, but agree on the linear relation between DGT and $0\nu\beta\beta$ NMEs.
As the EFT can describe DGT NMEs, with LECs constrained by data from two single GT transitions, the correlation between DGT and $0\nu\beta\beta$ allows the EFT for heavy nuclei to access $0\nu\beta\beta$. This way, we can obtain EFT results for $0\nu\beta\beta$ NMEs with corresponding theoretical uncertainties.

This work is organized as follows. In Sec.~\ref{EFT_GT} we introduce the basics of the EFT for heavy nuclei, followed by a discussion of the GT transition operator and the uncertainties associated with it. In Sec.~\ref{DGT} the method to determine the NMEs for $0\nu\beta\beta$ decay is presented, which includes the calculation of DGT transitions within the framework of the EFT and their correlation with $0\nu\beta\beta$ decays. The resulting predictions of the EFT for $0\nu\beta\beta$ NMEs are compared to results from NSM, EDF, IBM, the quasiparticle random-phase approximation (QRPA), as well as \textit{ab initio} calculations. Finally, we conclude with a summary in Sec.~\ref{summary}.

\section{Effective field theory for Gamow-Teller \texorpdfstring{$\boldsymbol{\beta}$}{} decay}
\label{EFT_GT}

The EFT used here for DGT transitions and to estimate the NME for $0\nu\beta\beta$ decay follows previous work on $\beta$ decays~\cite{Coe18Gamow-Teller}. In this framework, the low-energy properties of even-even, odd-odd, and odd-mass nuclei are described in terms of collective excitations coupled to a neutron, neutron hole, proton, or proton hole~\cite{Co15ETvibuncqant,Co16nETvoo}. The effective operators associated with the observables of interest are systematically constructed in terms of the creation and annihilation operators of the collective mode and the fermions. A power-counting scheme suppresses terms in an effective operator containing a larger number of collective creation and annihilation operators. Thus, the building blocks from which the effective operators are constructed are
(\textit{i}) the bosonic creation and annihilation operators of the collective mode, $d^{\dagger}_{\mu}$ and $d_{\mu}$, representing quadrupole excitations of the even-even reference state $|0\rangle$,
and (\textit{ii}) the fermionic creation and annihilation operators $n^{\dagger}_{\mu}$ and $n_{\mu}$ ($p^{\dagger}_{\mu}$ and $p_{\mu}$), which create a neutron or a neutron hole (proton or proton hole) depending on the nucleus under investigation (the annihilation operators annihilate a fermion or fermion hole).
These operators fulfill the usual commutator and anticommutator relations:
\begin{equation}
\label{commutator}
[d_{\mu},d_{\nu}^{\dagger}]=\delta_{\mu\nu} \,, \quad
\{n_{\mu},n_{\nu}^{\dagger}\}=\delta_{\mu\nu} \,, \quad
\{p_{\mu},p_{\nu}^{\dagger}\}=\delta_{\mu\nu} \,.
\end{equation}
While the creation operators transform as the components of a spherical tensor under rotations~\cite{Ro10FundNucMod}, the annihilation operators do not. To simplify the construction of spherical-tensor operators, spherical annihilation operators are defined with components
\begin{align}\label{Eqannihilationoperator}
\tilde{a}_\mu=(-1)^{j_a+\mu}a_{-\mu},
\end{align}
where $a$ can be $d$, $n$, or $p$. For the fermion annihilation tensors, $j_a$ is the total angular momentum of the effective orbital which the fermion or fermion hole occupies. The angular momentum of quadrupole phonons is $j_d=2$.

Low-lying states in the nuclei of interest are described as excitations of the even-even reference state by successive application of creation operators. For example, the low-lying spectrum of even-even spherical nuclei is described as one- and two-phonon excitations of the reference state
\begin{equation}
\label{evenvenexcited}
|2M1\rangle=d^{\dagger}_{M}|0\rangle\,, \quad |JM2\rangle=\sqrt{\frac{1}{2}}\,(d^{\dagger}\otimes d^{\dagger})^{(J)}_{M} |0\rangle \,.
\end{equation}
The labels $J$, $M$, and $\mathcal{N}$ in $|JM\mathcal{N}\rangle$ specify the total angular momentum of the state, its total-angular-momentum projection, and the number of phonons.
The ground state of an odd-mass nucleus adjacent to the reference state is described as a fermion excitation coupled to the core, e.g.,
\begin{equation}
\label{oddmassground}
|j_\mathrm{f} M 0\rangle=f^\dagger_{M}|0\rangle \,,
\end{equation}
where $\mathrm{f}=\mathrm{p},\mathrm{n}$, depending on the unpaired nucleon or nucleon  hole of the nucleus under consideration. For example, $^{93}$Y can be described by coupling a proton to a $^{92}$Sr core or by coupling a proton hole to a $^{94}$Zr core. The total angular momentum of the odd-mass ground state is thus equal to the total angular momentum $j_\mathrm{f}$ of the effective orbital of the unpaired nucleon or nucleon-hole.

In this framework, states in an odd-odd nucleus can be described by coupling a proton (or proton hole) and a neutron (or neutron hole) creation operator to the even-even reference state
\begin{equation}
\label{oddoddground}
|JM0;j_\mathrm{p},j_\mathrm{n}\rangle=(n^{\dagger}\otimes p^{\dagger})^{(J)}_{M}|0\rangle \,.
\end{equation}
For example, $^{74}$As can be described as coupling a neutron and proton hole to $^{74}$Ge or as coupling a neutron hole and proton to $^{74}$Se.

\subsection{Effective Gamow-Teller operator}

The effective GT operator can be constructed as the most general spherical tensor of rank one resulting from coupling bosonic and fermionic creation and annihilation operators. At lowest order in the number of boson operators it takes the form~\cite{Coe18Gamow-Teller}
\begin{align}
\hat{O}_{\rm GT} = & C_{\beta}
\left(\tilde{p} \otimes \tilde{n} \right)^{(1)} \nonumber \\
& + \sum\limits_{\ell} C_{\beta \ell} \left[
\left(d^{\dagger} + \tilde{d}\right) \otimes
\left( \tilde{p} \otimes \tilde{n} \right)^{(\ell)} \right]^{(1)} \nonumber \\
& + \sum\limits_{L\ell} C_{\beta L\ell} \left[
\left(d^{\dagger} \otimes d^{\dagger} +
\tilde{d}\otimes\tilde{d}\right)^{(L)} \otimes
\left( \tilde{p} \otimes \tilde{n} \right)^{(\ell)} \right]^{(1)} \nonumber \\[1mm]
& + {\rm H.c.} \,,
\label{GTEFT}
\end{align}
where $C_{\beta}$, $C_{\beta \ell}$, and $C_{\beta L \ell}$ are LECs that must be determined by matching to data or other theoretical input. The first, second, and third terms in Eq.~(\ref{GTEFT}) couple states of the odd-odd and even-even nuclei with phonon-number differences of 0, 1, and 2, respectively. Thus, they are involved in the description of $\beta$ decays from the intermediate odd-odd nucleus, as well as charge-exchange reactions between the same systems. Data on these transitions provide the necessary input to determine the LECs of the effective GT operator in the EFT. Note that if $C_\beta$ contributes to the transition, which is always the case in this work, this is the leading order (LO) contribution, and the other $C_{\beta \ell}$ and $C_{\beta L \ell}$ parts constitute higher-order corrections. 

The $ft$ value of GT decay is given by
\begin{equation}
\label{ft}
\left(ft\right)_{if}=\frac{\kappa}{g^2_A}\frac{2J_i+1}{\left|M_\mathrm{GT}(J_i\rightarrow J_f)\right|^2} \,,
\end{equation}
where $f$ is a phase-space factor, $\kappa=6147$\,s the $\beta$ decay constant, $g_A=1.27$ the axial-vector coupling, $J_i$ ($J_f$) the total angular momentum of the initial (final) state, and $M_\text{GT}$ the NME of the GT operator between the initial and final nucleus. Moreover, the relation between the GT strength $S_\pm(i \to f)$ measured in charge-exchange reactions and the GT NME is given by $S_\pm(i \to f)=|M_\text{GT}|^2$. For details in the EFT see Ref.~\cite{Coe18Gamow-Teller}.

The power-counting scheme suggests that contributions from the different terms in Eq.~(\ref{GTEFT}) scale according to the ratio of the typical energy to the breakdown scale $\Lambda$ of the EFT~\cite{Co15ETvibuncqant, Co16nETvoo, Coe18Gamow-Teller}. In particular, the NMEs of an operator containing $n$ quadrupole operators scale as
\begin{equation}
\langle d^{n} \rangle \sim
\left(\frac{\Lambda}{\omega}\right)^{n/2}\,,
\label{power}
\end{equation}
where $\omega$ is the energy scale of the collective mode. For the different terms in the GT operator, this leads to
\begin{equation}
C_{\beta}\langle d^{0} \rangle \sim
C_{\beta \ell} \langle d^{1} \rangle
\quad {\rm or} \quad
\frac{C_{\beta \ell}}{C_{\beta}} \sim
0.58 \pm 0.38
\label{scale}
\end{equation}
and
\begin{equation}
C_{\beta} \langle d^{0} \rangle \sim
C_{\beta L\ell} \langle d^{2} \rangle
\quad {\rm or} \quad
\frac{C_{\beta L\ell}}{C_{\beta}} \sim
0.33 \pm 0.22\,,
\label{scale2}
\end{equation}
assuming a breakdown scale at the three phonon level, $\Lambda=3\omega$, due to neglected physics such as pairing effects, which would need to be taken into account at this energy scale~\cite{Co15ETvibuncqant}. $\Lambda=3\omega$ fixes the ratio of the LECs. The above uncertainties associated to these ratios are 68$\%$ degree of belief and have been estimated based on the expectation for the LECs to be natural using prior distributions of the form
\begin{align}
\mathrm{pr}(C|c)&=\frac{1}{\sqrt{2\pi}c}e^{-\frac{1}{2}\left(\frac{C-1}{s c}\right)^2} \,,\\
\mathrm{pr}(c)&=\frac{1}{\sqrt{2\pi}\sigma c}e^{-\frac{1}{2}\left(\frac{\mathrm{ln} c}{\sigma}\right)^2 } \,,
\end{align}
with $\sigma=\mathrm{ln}(3/2)$ and $s=0.65$. For a LEC, the interval $1-s\leqslant C \leqslant 1+s$ of the prior distribution yields 68$\%$ degree of belief. For electromagnetic transitions it was shown in Ref.~\cite{Co15ETvibuncqant} that the cumulative distribution of next-to-leading-order (NLO) LECs is well approximated by the above priors. For the effective GT operator, a similar strategy was successfully applied to $\beta$ and two-neutrino $\beta\beta$/ECEC decays~\cite{Coe18Gamow-Teller, Perez:2018cly}, showing
good agreement with experiment within uncertainties.

\subsection{Uncertainty estimate for decays to ground state}

Within the EFT, the sources for uncertainty in the NME are (\textit{i}) omitted terms in the effective GT operator, Eq.~(\ref{GTEFT}), which involve two or more boson operators (for them to couple odd-odd and even-even ground states); these are expected to scale as $(\omega/\Lambda)C_{\beta}$; (\textit{ii}) omitted NLO corrections to odd-odd states due to terms in the Hamiltonian mixing states with phonon-number differences of one, expected to scale as $\sqrt{\omega/\Lambda}|JM;j_\mathrm{p},j_\mathrm{n}\rangle$. These would then be able to couple with corrections to the GT operator that scale only with one or more boson operators. (For information on the Hamiltonian of the EFT, which shows good agreement with experiment within the estimated uncertainties, see Refs.~\cite{Co15ETvibuncqant,Co16nETvoo}.) Overall, we therefore expect the omitted correction to the NME of the GT decay to even-even ground states due to both sources to scale as
\begin{equation}
\Delta M_{\rm GT}\left( 1^{+} \rightarrow 0^{+}_\mathrm{gs} \right)
\overset{\rm EFT}{\sim} \frac{\omega}{\Lambda} \,
M_{\rm GT}\left( 1^{+} \rightarrow 0^{+}_\mathrm{gs} \right)\,.
\label{Deltagstogs}
\end{equation}

\section{Double-Gamow-Teller transitions and neutrinoless double-\texorpdfstring{$\boldsymbol{\beta}$}{} decays}
\label{DGT}

\subsection{Double-Gamow-Teller transitions}
\label{DGT_EFT}

The goal of this work is to predict NMEs governing the $0\nu\beta\beta$ decay of medium-mass to heavy nuclei within the EFT presented in the previous section. The $0\nu\beta\beta$ decay mode is more challenging to study compared to two-neutrino $\beta\beta$/ECEC decays, as the lack of experimental data prevents the determination of the LECs in the $0\nu\beta\beta$ decay operator. Thus an alternative method to access $0\nu\beta\beta$ NMEs is required. Recently, it was shown that DGT NMEs correlate with $0\nu\beta\beta$ NMEs~\cite{Shimizu:2017qcy}; for details see also Refs.~\cite{Me_0nbb1,Me_0nbb2}. Using this correlation, we can thus estimate $0\nu\beta\beta$ NMEs from EFT calculations of DGT NMEs, where the LECs can be obtained from experimental data on GT transitions.

In the EFT, the DGT operator can be written as the coupling of two GT operators, Eq.~(\ref{GTEFT}), to total angular momentum zero. The lowest order contribution coupling the ground state of the initial and final nuclei is given by
\begin{align}
\hat{O}_{\rm DGT} =&  
\left( \hat{O}_{\rm GT} \otimes \hat{O}_{\rm GT} \right)^{(0)} \nonumber \\
=& C_{\beta_1}C_{\beta_2}
\left( \left( \tilde{p} \otimes \tilde{n} \right)^{(1)} \otimes
\left( \tilde{p} \otimes \tilde{n} \right)^{(1)} \right)^{(0)}
+ {\rm H.c.} \nonumber \\
&+ \ldots\,,
\label{eDGT}
\end{align}
where the LECs $(C_{\beta_1}$ and $C_{\beta_2})$ are extracted from the $\log(ft)$ value of $\beta$ decays of the intermediate odd-odd nucleus or the zero-angle cross sections of the corresponding charge-exchange reactions. 
The dots in Eq.~(\ref{eDGT}) denote higher-order terms, which we do not consider but enter the EFT uncertainty estimates.

For the ground state of the initial nucleus, we assume that this can be written at LO within the EFT as a multifermion excitation of the reference state, i.e.,
\begin{equation}
|0^{+}_\mathrm{gs} \rangle = \frac{1}{2}
\left( n^{\dagger}\otimes n^{\dagger}\right)^{(0)}
\left( p^{\dagger}\otimes p^{\dagger}\right)^{(0)} |0 \rangle\,,
\label{initial}
\end{equation}
where the total angular momenta and parities of the effective orbitals, in which the neutrons and protons lie, are inferred from the low-lying states of the odd-mass nuclei adjacent to the intermediate nucleus in the DGT transition. At LO the initial ground state is built by like fermions coupled to spin zero. 
We have checked that the inclusion of couplings to higher $J$ does not significantly change our results, as it only enlarges the uncertainties to some extent where coupling to higher $J$ is allowed at LO.
The dominance of $J=0$ pairs parallels the dominance of $s$ bosons in the IBM~\cite{skouras1990,bonatsos1991}.

Thus, the NME of the DGT operator Eq.~(\ref{eDGT}) between even-even ground states results in
\begin{equation}
\langle 0^{+}_{\mathrm{gs},f} || \hat{O}_{\rm DGT} || 0^{+}_{\mathrm{gs},i}\rangle
= \sqrt{\frac{4}{3(2j_\mathrm{n}+1)(2j_\mathrm{p}+1)}}
{C}_{\beta_1}{C}_{\beta_2} \,,
\label{eq:0to0nless}
\end{equation}
taking Eq.~(\ref{initial}) as a good approximation for the ground state of the initial nucleus. 
The LECs of the effective DGT operator, $C_{\beta_1}$ and $C_{\beta_2}$, are fitted to GT data, which only include the lowest lying $1^+_1$ in the intermediate nucleus. However in the DGT transition, the virtual single GT transition from the intermediate nucleus to the initial or final nucleus could also include higher-lying $1^+_{n+1}$ states of the intermediate nucleus. The effective DGT operator would then require NLO and higher order terms with two and more additional quadrupole operators to account for these excited $1^+_{n+1}$ intermediate states. According to the power counting, Eq.~(\ref{power}), the LECs of the GT transition to the excited $1^+_{n+1}$ intermediate states scale as
\begin{equation}
    C_{\beta,1^+_{n+1}} \sim \frac{\langle d^0\rangle}{\langle d^n\rangle}C_{\beta,1^+_1}\sim  \left(\frac{\omega}{\Lambda}\right)^{n/2} C_{\beta,1^+_1} \,.
\end{equation}
Thus, the relative uncertainty in the NME of Eq.~(\ref{eq:0to0nless}) can be naively and conservatively estimated by summing over all neglected contributions with the same phase
\begin{equation}\label{delta}
\Delta \langle 0^{+}_{\mathrm{gs},f} || \hat{O}_{\rm DGT} || 0^{+}_{\mathrm{gs},i}\rangle \sim \sum_{n=1}
\left( \frac{\omega}{\Lambda} \right)^{n} = \frac{1}{2} \,.
\end{equation}
Note that a similar scaling correction would arise when including NLO and higher order corrections from the Hamiltonian, as discussed for GT transitions above. We adopt this relative uncertainty, Eq.~(\ref{delta}), for the DGT NME, Eq.~(\ref{eq:0to0nless}), with one orbital combination $j_\mathrm{n}, j_\mathrm{p}$. In the case when several $j_\mathrm{n}, j_\mathrm{p}$ are possible (see below), we average the individual DGT NMEs and adopt their combined uncertainty range. As an example for a DGT NME including several orbital combinations, the normalized DGT NME of $^{96}$Zr is shown in Fig~\ref{figunc}. Finally, we emphasize that the EFT NMEs match the LECs directly to data, which thus encodes information on the quenching of $g_{A}$ through correlations or two-body-current contributions in more complex calculations~\cite{Gysbers:2019uyb}.

\begin{figure}[t]
\centering
\includegraphics[width=0.8\columnwidth]{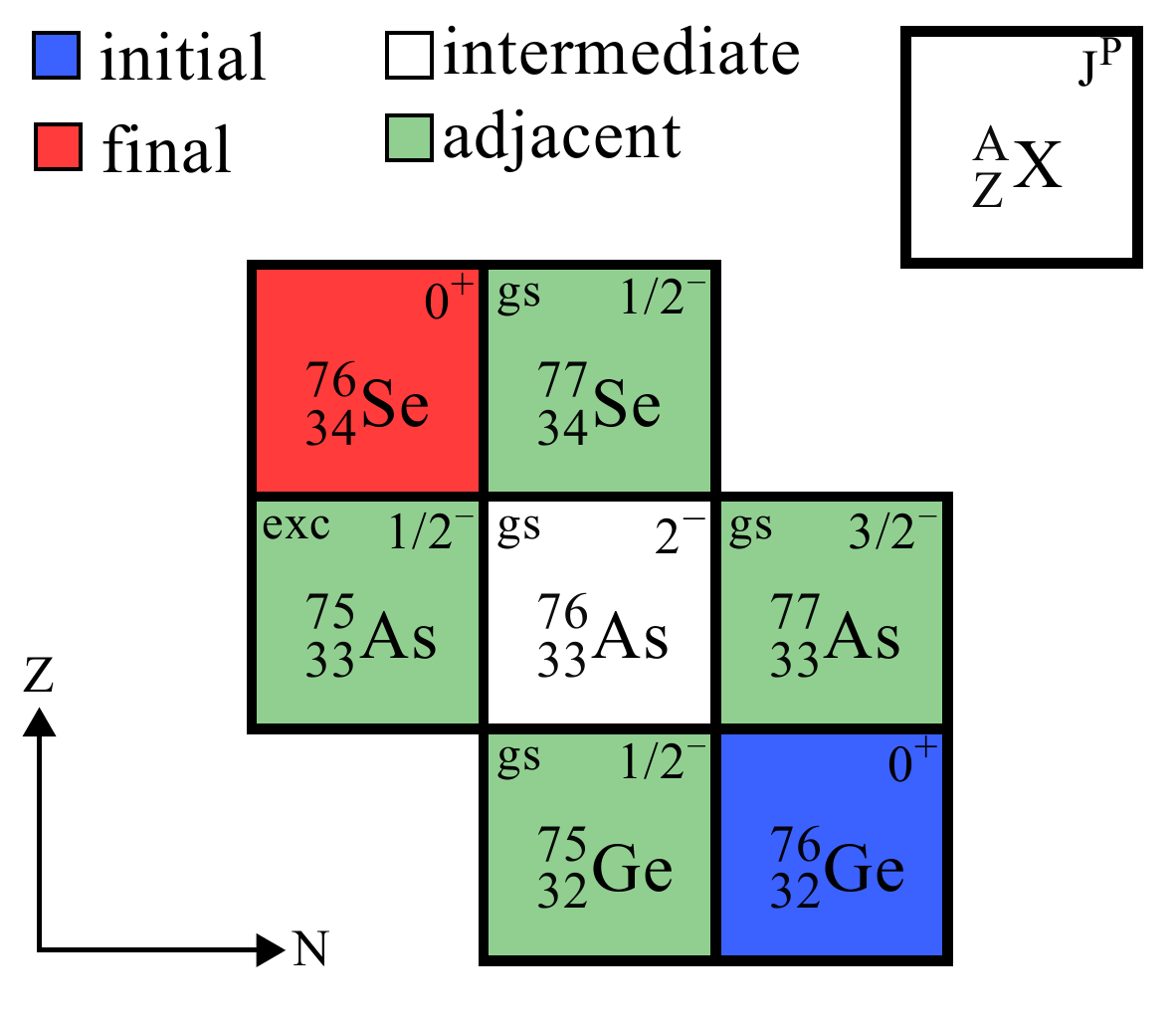}
\caption{Illustration of the initial (blue), intermediate (white), and final (red) nucleus for the $0\nu\beta\beta$ decay and DGT transition of $^{76}$Ge. The relevant adjacent nuclei are shown in green. For each nucleus the upper right corner shows the spin and parity of the ground (gs) or excited (exc) state used to derive orbital combinations $j_\mathrm{n}, j_\mathrm{p}$ of the DGT transition.}
\label{figjnjp}
\end{figure}

To calculate the DGT NME, Eq.~(\ref{eq:0to0nless}), first the total angular momenta and parities of the neutron and proton in the odd-odd intermediate nucleus must be determined. For the initial nucleus to undergo a DGT transition, it is required that they are able to couple to a $1^+$ state. As an example, Fig.~\ref{figjnjp} shows possible $j_\mathrm{n}$ and $j_\mathrm{p}$ combinations for the case of the DGT transition from $^{76}$Ge to $^{76}$Se. If a proton (neutron) is added to or removed from the odd-odd intermediate nucleus, one obtains an odd-mass nucleus with an odd neutron (proton). The total angular momenta and parities of the effective orbitals in which these neutrons and protons lie can thus be inferred from the spectra of the adjacent odd-mass nuclei $^{77}$Se, $^{75}$Ge, $^{77}$As, and $^{75}$As. As shown in Fig.~\ref{figjnjp}, the low-lying spectra of these adjacent nuclei suggest more than one orbital combination. While the ground states of $^{75}$Ge and $^{77}$As suggest $j_\mathrm{n}=\tfrac{1}{2}^-$ and $j_\mathrm{p}=\tfrac{3}{2}^-$, the excited $\tfrac{1}{2}^-$ state in  $^{75}$As also suggests $j_\mathrm{p}=\tfrac{1}{2}^-$. In this case, when an excited state is used to determine $j_\mathrm{n/p}^P$, only states with lifetimes $t_{1/2}\geqslant0.1$\,ns are considered to exclude collective excitations. With this selection we want to rule out states that cannot be described in the EFT as a single-nucleon excitation. In addition, we only consider excited states below 700\,keV to determine $j_\mathrm{n}^P$, $j_\mathrm{p}^P$ of the effective nucleon orbitals, as high-energy single-particle states in the adjacent nuclei are unlikely to contribute to the $1^+$ ground state of the intermediate odd-odd nucleus in the DGT transition.

In Table~\ref{tab:spinparity} the total angular momenta and parity of all considered states are listed, as well as their energy in the spectrum of the adjacent odd-mass nucleus. As can be seen, for some nuclei there is more than one combination possible to couple to a $1^+$ state for the odd-odd intermediate nucleus, e.g., as shown in Fig.~\ref{figjnjp} for the DGT transition from $^{76}$Ge.

For $^{96}$Zr, no states in the spectra of the odd-mass nuclei were able to satisfy the conditions stated above. For $^{116}$Cd, only one orbital combination satisfies the conditions, but one of the orbitals lies very close to 700\,keV (see Table~\ref{tab:neglected}). In this case, we chose to neglect this orbital combination. As discussed below, this leads to a larger uncertainty. In both cases, we thus assume neutron and proton total angular momenta ranging from $\tfrac{1}{2}$ to $\tfrac{9}{2}$ to predict their DGT NMEs, since these are the most frequent, as can be seen in Table~\ref{tab:spinparity}.

\begin{figure}[t]
\centering
\includegraphics[width=\columnwidth]{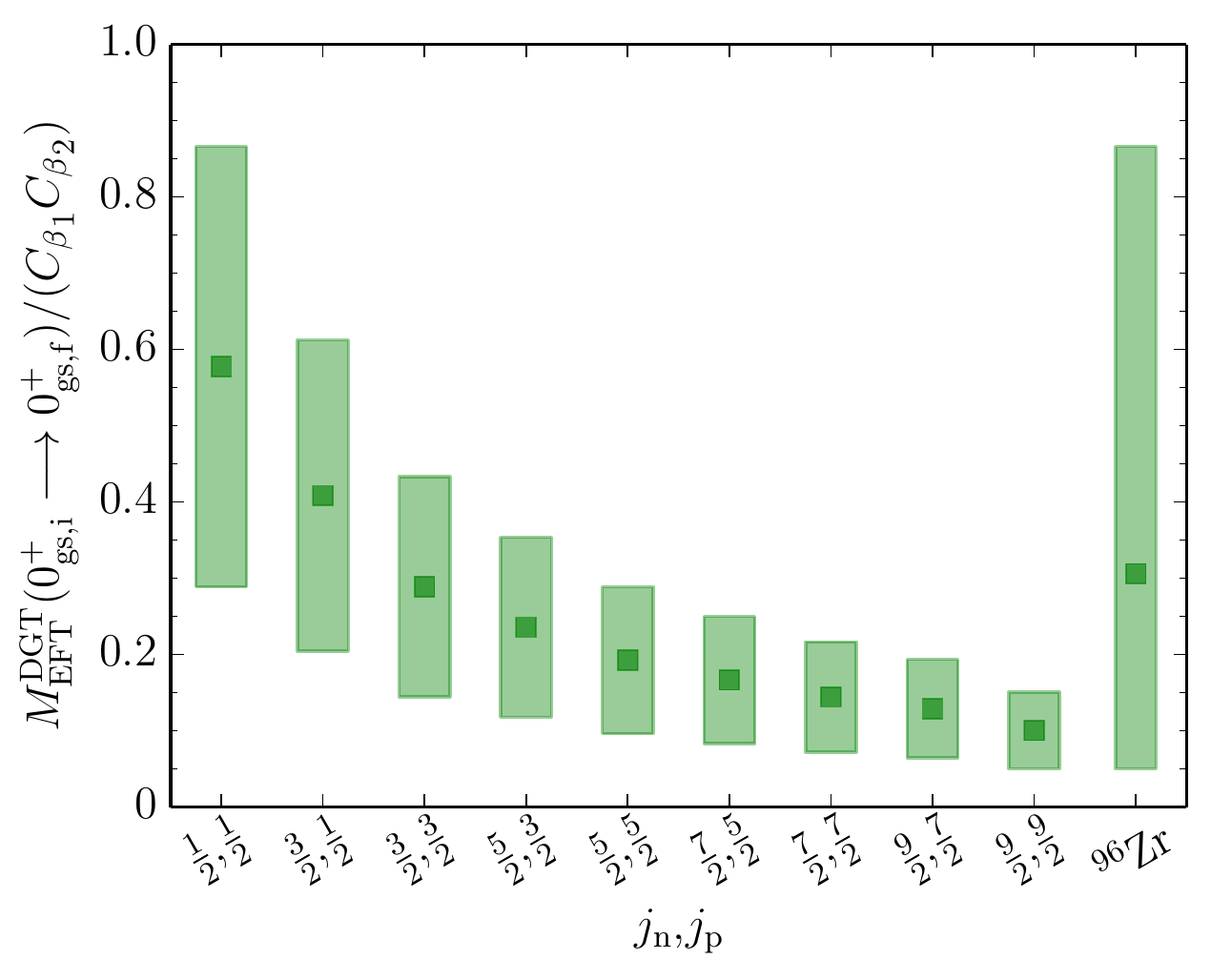}
\caption{Double-Gamow-Teller NMEs in the EFT for different combinations of neutron and proton orbitals $j_\mathrm{n}, j_\mathrm{p}$, normalized by the product of the LECs $C_{\beta_1} C_{\beta_2}$. The squares and bars give the central value and EFT truncation uncertainty, respectively. As an example, the normalized $^{96}$Zr DGT NME on the right covers all orbital combinations $\tfrac{1}{2}\leqslant j_\mathrm{n}, j_\mathrm{p}\leqslant \tfrac{9}{2}$.}
\label{figunc}
\end{figure}

Figure~\ref{figunc} shows the normalized DGT NMEs for different combinations of neutron and proton orbitals and their corresponding uncertainties. Since there is more than one combination of neutron and proton orbitals possible, this introduces another source of uncertainty to the DGT NME. In Fig.~\ref{figunc} only one bar is plotted for orbital combinations where the total angular momenta of the neutron and proton orbitals are unequal. In case both combinations for $j_\mathrm{n}\neq j_\mathrm{p}$ are allowed, we count this twice for the average of the normalized DGT NME over all orbital combinations. As discussed above and shown in Fig.~\ref{figunc}, all possible combinations of $\tfrac{1}{2}\leqslant j_\mathrm{n}, j_\mathrm{p} \leqslant \tfrac{9}{2}$ are included into the prediction of $^{96}$Zr and $^{116}$Cd. The uncertainty is taken to be the combined uncertainty range of the included normalized DGT NMEs.

\begin{table*}[t]
    \centering
    \caption{For DGT and $0\nu\beta\beta$ decay of the initial nucleus (column 1), we give the spin-parity (column 2 and 7) and energy of states (column 3, 5, 8 and 10) of adjacent (adj.) nuclei (column 4, 6, 9 and 11) that differ by one neutron (leading to $j_\mathrm{n}^P$) or one proton ($j_\mathrm{p}^P$) from the initial or final nuclei of the decay. Data were taken from Refs.~\cite{NDS47, NDS49, NDS65, NDS69, NDS71, NDS75, NDSA77, NDS79, NDS81, NDS83, NDS95, NDS97, NDS99, NDS103, NDS105, NDS107, NDS109, NDS111, NDS113, NDS115, NDS123, NDS125, NDS127, NDS129, NDS131, NDS135, NDS137, NDS149, NDS151} and for $^{63}$Ni, $^{63}$Cu, $^{77}$Se, $^{77}$As, $^{101}$Ru, $^{101}$Tc, $^{108}$Cd, $^{117}$Sn, and $^{117}$In from the ENSDF database \cite{NDSlist}. The allowed $j_\mathrm{n}^P, j_\mathrm{p}^P$ combinations must couple to $1^+$, and we consider only states with lifetimes $t_{1/2}\geqslant0.1$\,ns to exclude collective excitations. Additional, neglected states are discussed and listed in Appendix~\ref{neglected}.\\} 
    \begin{tabular}{c|c r l r l| c r l r l}\hline\hline
        Initial & $j_\mathrm{n}^P$ & $E$\,[keV] & \multicolumn{1}{c}{adj.} & $E$\,[keV] & \multicolumn{1}{c|}{adj.} & $j_\mathrm{p}^P$ & $E$\,[keV] & \multicolumn{1}{c}{adj.}& $E$\,[keV] & \multicolumn{1}{c}{adj.}\\\hline\hline&&&&&&\\[-4mm]
             $^{~48}$Ca & ${7}/{2}^-$ &0.{\color{white}00} & $^{~49}$Ti & 0.{\color{white}00} & $^{~47}$Ca & ${7}/{2}^-$ & 0.{\color{white}00} & $^{~49}$Sc& 0.{\color{white}00} & $^{~47}$Sc\\&&&&&&\\[-4mm]\hline&&&&&&\\[-4mm]
             \multirow{3}{*}{$^{~64}$Zn}& $1/2^-$ & $53.93$ & & 0.{\color{white}00} &  &\multirow{3}{*}{$3/2^-$}& \multirow{3}{*}{0.{\color{white}00}} & \multirow{3}{*}{$^{~65}$Cu}& \multirow{3}{*}{0.{\color{white}00}} & \multirow{3}{*}{$^{~63}$Cu}\\
             & $3/2^-$ & $115.13$ & $^{~65}$Zn& \footnotemark[1]$155.55$ & $^{~63}$Ni& &&&\\
             & $5/2^-$ & 0.{\color{white}00} & & 87.15 & & &&&\\&&&&&&\\[-4mm]\hline&&&&&&\\[-4mm]
             \multirow{2}{*}{$^{~70}$Zn}& {${1}/{2}^-$}& 0.{\color{white}00} & \multirow{2}{*}{$^{~71}$Ge}& 0.{\color{white}00} & \multirow{2}{*}{$^{~69}$Zn}& \multirow{2}{*}{${3}/{2}^-$}& \multirow{2}{*}{0.{\color{white}00}} & \multirow{2}{*}{$^{~71}$Ga}&  \multirow{2}{*}{0.{\color{white}00}} & \multirow{2}{*}{$^{~69}$Ga}\\
             &{${5}/{2}^-$}& 174.94 &  & \footnotemark[1]531.30 & & & &&&\\&&&&&&\\[-4mm]\hline&&&&&&\\[-4mm]
             \multirow{3}{*}{$^{~76}$Ge}&\multirow{2}{*}{${1}/{2}^-$}& \multirow{2}{*}{0.{\color{white}00}} & & \multirow{2}{*}{0.{\color{white}00}} & & {${3}/{2}^-$}& 0.{\color{white}00} & & 0.{\color{white}00} & \\
             &\multirow{2}{*}{{${5}/{2}^\mp$}}& \multirow{2}{*}{249.79} & $^{~77}$Se& \multirow{2}{*}{192.19} & $^{~75}$Ge&{${1}/{2}^-$}& \footnotemark[1]503.88 & $^{~77}$As&  198.61 & $^{~75}$As\\
             &&&&&&{${5}/{2}^\mp$}& 264.43 &  & 400.66 & \\&&&&&&\\[-4mm]\hline&&&&&&\\[-4mm]
             \multirow{2}{*}{$^{~80}$Se}& {${1}/{2}^-$}& 190.64 & \multirow{2}{*}{$^{~81}$Kr}&95.77 & \multirow{2}{*}{$^{~79}$Se}& {${3}/{2}^-$}& 0.{\color{white}00} & \multirow{2}{*}{$^{~81}$Br}& 0.{\color{white}00} & \multirow{2}{*}{$^{~79}$Br}\\
             &{${9}/{2}^+$}& 49.57 & & \footnotemark[1]136.97 &  &{${9}/{2}^+$}& 536.20 &  & 207.61 & \\&&&&&&\\[-4mm]\hline&&&&&&\\[-4mm]
             \multirow{2}{*}{$^{~82}$Se}& {${1}/{2}^-$}& 41.56  & \multirow{2}{*}{$^{~83}$Kr}&0.{\color{white}00} & \multirow{2}{*}{$^{~81}$Se}& {${3}/{2}^-$}& 0.{\color{white}00} & \multirow{2}{*}{$^{~83}$Br}& 0.{\color{white}00} & \multirow{2}{*}{$^{~81}$Br}\\
             &{${9}/{2}^+$}& 0.{\color{white}00} & & \footnotemark[1]294.30 &  &{${9}/{2}^+$}& 1092.10 &  & 536.20 & \\&&&&&&\\[-4mm]\hline&&&&&&\\[-4mm]
             \multirow{3}{*}{$^{100}$Mo}& {${7}/{2}^+$}& \footnotemark[1]842.76 &\multirow{3}{*}{$^{101}$Ru}  &235.51 & \multirow{3}{*}{$^{~99}$Mo} & {${9}/{2}^+$}& 0.{\color{white}00} & \multirow{3}{*}{$^{101}$Tc}&0.{\color{white}00} & \multirow{3}{*}{$^{~99}$Tc}\\
             &{${5}/{2}^+$}& 0.{\color{white}00} & & 97.79 && {${7}/{2}^+$}& 9.32 &&  140.51\\
             &{${3}/{2}^+$}& 127.23 && 351.22 & &{${5}/{2}^+$}& 15.60 & &181.09\\&&&&&&\\[-4mm]\hline&&&&&&\\[-4mm]
             {$^{104}$Ru}&{${5}/{2}^+$}& 0.{\color{white}00} & $^{105}$Pd & \footnotemark[1]2.81 & $^{103}$Ru &{${7}/{2}^+$}& 0.{\color{white}00} & $^{105}$Rh& 39.75 & $^{103}$Rh\\&&&&&&\\[-4mm]\hline&&&&&&\\[-4mm]
             \multirow{2}{*}{$^{106}$Cd}& {${5}/{2}^+$}& 0.{\color{white}00} & \multirow{2}{*}{$^{107}$Cd}& 0.{\color{white}00} & \multirow{2}{*}{$^{105}$Pd}&{${7}/{2}^+$}& 93.13 & \multirow{2}{*}{$^{107}$Ag} & 25.47 & \multirow{2}{*}{$^{105}$Ag}\\
             &{${7}/{2}^+$}& 204.98 & & \footnotemark[1]785.00&&{${9}/{2}^+$} &125.59&&53.14&\\&&&&&&\\[-4mm]\hline&&&&&&\\[-4mm]
             {$^{108}$Cd}& {${5}/{2}^+$}& 0.{\color{white}00} & $^{109}$Cd &0.{\color{white}00} & $^{107}$Pd &{${7}/{2}^+$}& 88.03 & $^{109}$Ag & 93.13 & $^{107}$Ag \\&&&&&&\\[-4mm]\hline&&&&&&\\[-4mm]
             \multirow{2}{*}{$^{110}$Pd}&{${5}/{2}^+$}& 245.39 & \multirow{2}{*}{$^{111}$Cd}& 0.{\color{white}00} & \multirow{2}{*}{$^{109}$Pd}&{${7}/{2}^+$}& 59.82 & \multirow{2}{*}{$^{111}$Ag}& 88.03 & \multirow{2}{*}{$^{109}$Ag}\\
             &{${7}/{2}^+$}& 416.72 &  & \footnotemark[1]645.96 &  &{${9}/{2}^+$}& 130.28 &  & 132.76 & \\&&&&&&\\[-4mm]\hline&&&&&&\\[-4mm]
             {$^{112}$Sn}& {${7}/{2}^+$}& 77.39 & $^{113}$Sn  & 416.72 & $^{111}$Cd &{${9}/{2}^+$}& 0.{\color{white}00} & $^{113}$In& 0.{\color{white}00} & $^{111}$In \\&&&&&&\\[-4mm]\hline&&&&&&\\[-4mm]
             {$^{114}$Cd}& {${7}/{2}^+$}& 612.81 & $^{115}$Sn& \footnotemark[1]458.63 & $^{113}$Cd&{${9}/{2}^+$}& 0.{\color{white}00} & $^{115}$In& 0.{\color{white}00} & $^{113}$In \\&&&&&&\\[-4mm]\hline&&&&&&\\[-4mm]
             \multirow{3}{*}{$^{124}$Xe} & \multirow{2}{*}{${7}/{2}^+$}& \multirow{2}{*}{295.89} & \multirow{3}{*}{$^{125}$Xe}& \multirow{2}{*}{489.81} & \multirow{3}{*}{$^{123}$Te}&  {${5}/{2}^+$}& 0.{\color{white}00} & \multirow{3}{*}{$^{125}$I}& 0.{\color{white}00} & \multirow{3}{*}{$^{123}$I}\\
             &\multirow{2}{*}{${3}/{2}^+$}& \multirow{2}{*}{111.79} &&\multirow{2}{*}{159.02} &&{${7}/{2}^+$}& 113.54 &&\footnotemark[1]138.20 &\\
             &&&&&&{${9}/{2}^+$}& 935.70 &&641.25 &\\&&&&&&\\[-4mm]\hline&&&&&&\\[-4mm]
             \multirow{2}{*}{$^{128}$Te}& {${3}/{2}^+$}& 39.58 & \multirow{2}{*}{$^{129}$Xe} & 0.{\color{white}00} & \multirow{2}{*}{$^{127}$Te}&{${5}/{2}^+$}& 27.79 & \multirow{2}{*}{$^{129}$I}& 0.{\color{white}00} & \multirow{2}{*}{$^{127}$I}\\
             &{${1}/{2}^+$}& 0.{\color{white}00} &&\footnotemark[1]61.16 &&{${3}/{2}^+$}& 278.38 &&202.86 &\\&&&&&&\\[-4mm]\hline&&&&&&\\[-4mm]
             \multirow{2}{*}{$^{130}$Te}& {${3}/{2}^+$}& 0.{\color{white}00} & \multirow{2}{*}{$^{131}$Xe} & 0.{\color{white}00} & \multirow{2}{*}{$^{129}$Te}&{${5}/{2}^+$}& 149.72 & \multirow{2}{*}{$^{131}$I} & 27.79 & \multirow{2}{*}{$^{129}$I}\\
             &{${1}/{2}^+$}& 80.19 &&\footnotemark[1]180.36 & &{${3}/{2}^+$}& \footnotemark[1]492.66 && 278.38 &\\&&&&&&\\[-4mm]\hline&&&&&&\\[-4mm]
             {$^{136}$Xe}& {${3}/{2}^+$}& 0.{\color{white}00} & $^{137}$Ba & 0.{\color{white}00} & $^{135}$Xe&{${5}/{2}^+$}& \footnotemark[1]455.49&$^{137}$Cs& 249.77 & $^{135}$Cs\\&&&&&&\\[-4mm]\hline&&&&&&\\[-4mm]
             \multirow{6}{*}{$^{150}$Nd}& \multirow{2}{*}{${5}/{2}^+$}& \multirow{2}{*}{167.75} & \multirow{6}{*}{$^{151}$Sm}& \multirow{2}{*}{\footnotemark[1]332.94} & \multirow{6}{*}{$^{149}$Nd}&{${3}/{2}^+$}& 255.69 & \multirow{6}{*}{$^{151}$Pm} & 188.63 & \multirow{6}{*}{$^{149}$Pm}\\
             &\multirow{2}{*}{${9}/{2}^+$}&\multirow{2}{*}{91.53}&&\multirow{2}{*}{270.86}&&{${5}/{2}^+$}& 0.{\color{white}00} &&114.31 &\\
             &&&&&&{${7}/{2}^+$}& \footnotemark[1]85.12&&0.{\color{white}00} &\\
             &{${5}/{2}^-$}& 0.{\color{white}00}&&0.{\color{white}00}&&\multirow{2}{*}{${7}/{2}^-$}&\multirow{2}{*}{\footnotemark[1]175.08}&&\multirow{2}{*}{270.17}&\\
             &{${9}/{2}^-$}& \footnotemark[1]175.38 &&220.71 & &\multirow{2}{*}{${11}/{2}^-$}& \multirow{2}{*}{\footnotemark[1]343.80}&&\multirow{2}{*}{240.21}\\
             &{${11}/{2}^-$}& 261.13&&\footnotemark[1]192.00&&&&&&\\\hline\hline
        \end{tabular}
\footnotetext[1]{No data on half-life available or only upper limit above 0.1\,ns.}
\label{tab:spinparity}
\end{table*}

For all considered DGT transitions, the spin and parity of the ground states of the nuclei involved are given in Appendix~\ref{ratio}, along with checks of the EFT for spherical cores through the $E_{4^+}/E_{2^+}$ ratio. Given the allowed orbital combinations discussed above, the LECs ${C}_{\beta_1}$ and ${C}_{\beta_2}$ are fit to experimental GT decays or GT strengths from charge-exchange reactions for the intermediate nuclei~\cite{NDS64, NDS70, NDS80, NDS100, NDS104, NDS106, NDSlist, NDS110, NDS112, NDS114, NDS116, NDS128, BGT48, BGT76Ge, BGT82Se, BGT96, BGT130, BGT136, BGT150}, except for $^{124}\text{Xe}$ where data systematics on neighboring nuclei are used following Ref.~\cite{Perez:2018cly}. In some cases only one $\log(ft)$ value or GT strength of the investigated transition was available to obtain the LEC, which is then used for both ${C}_{\beta_1}$ and ${C}_{\beta_2}$. The resulting ${C}_{\beta_1}$ and ${C}_{\beta_2}$ and the EFT DGT NMEs are listed in Table~\ref{tab:predictions}. For cases with more than one possible orbital combination, the results are obtained with the same procedure as described above for $^{96}$Zr.

We have neglected some orbital combinations fulfilling the conditions of a DGT transition, but not listed in Table~\ref{tab:spinparity}. We do not consider orbitals if there are no experimental data on the lifetimes of the corresponding excited states in adjacent odd-mass nuclei, because this prevents us from identifying the states as dominantly single-particle excitations. For example, in $^{81}$Kr the $\tfrac{5}{2}^-$ state at 456.74$\,$keV could couple to the $\tfrac{3}{2}^-$ state in $^{79,81}$Br, and its energy is comparable to the one of the $\tfrac{9}{2}^+$ states in $^{79,81}$Br. However, the lifetime of the $\tfrac{5}{2}^-$ state in $^{81}$Kr is not known. In addition, we disregard orbitals when the NSM suggests that the corresponding states are not dominant or not of single-particle character. The reasons for omitting possible orbital combinations are discussed in more detail in Appendix~\ref{neglected} (see Table~\ref{tab:neglected} there).

\subsection{Correlation between \texorpdfstring{$\boldsymbol{M^{\rm DGT}}$}{} and \texorpdfstring{$\boldsymbol{M^{0\nu\beta\beta}}$}{}}

\begin{figure}[t]
\centering
\includegraphics[width=\columnwidth]{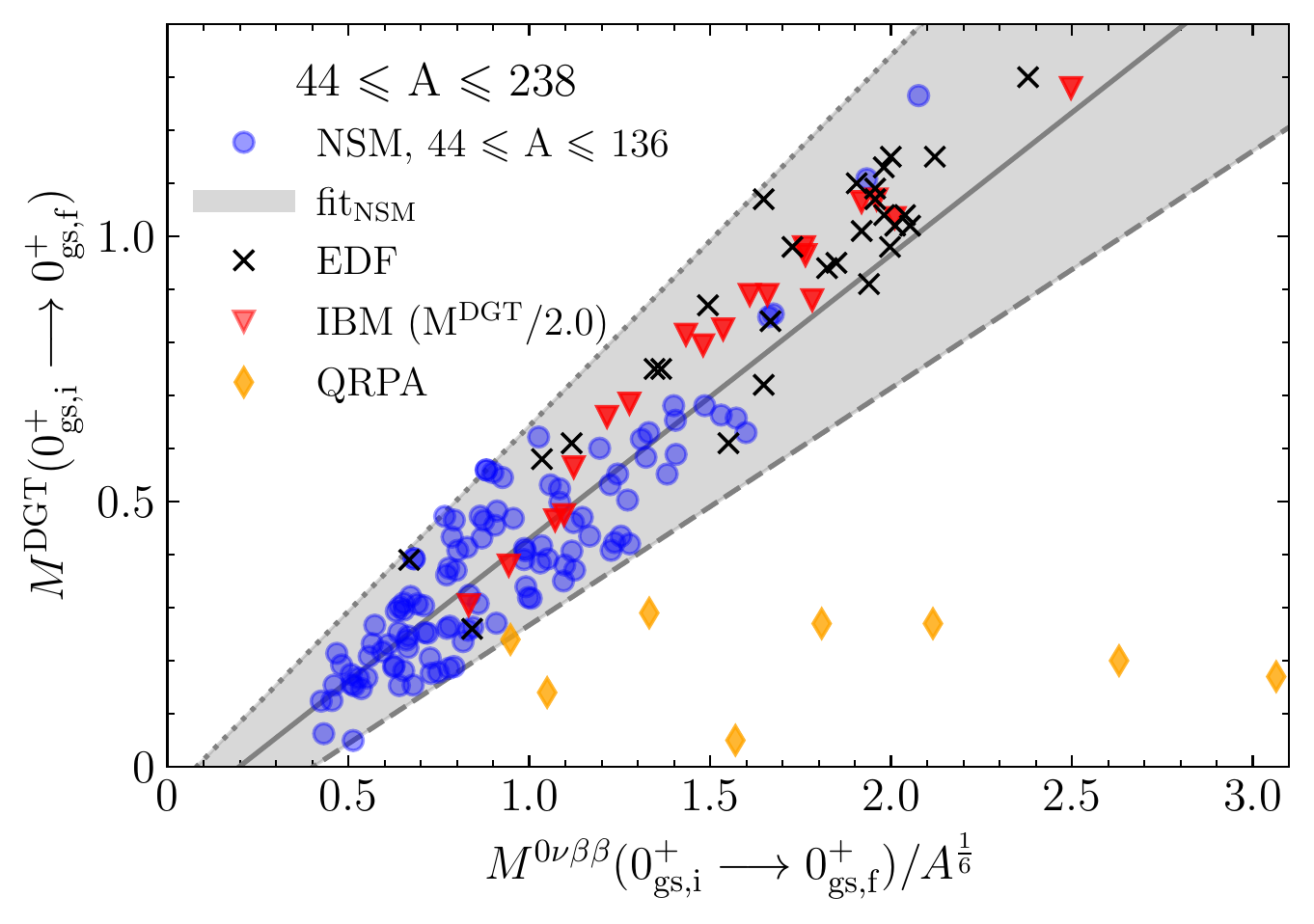}
\caption{Correlation between DGT and $0\nu\beta\beta$ NMEs, the latter divided by $A^{1/6}$.
The fit to NSM results (blue circles) gives a correlation coefficient $r_\mathrm{NSM}=0.90$. NMEs from EDF (black crosses), IBM (red stars), and QRPA calculations (orange diamonds) are also shown. Data were taken from Refs.~\cite{Shimizu:2017qcy,Me_0nbb1,Me_0nbb2}. The gray dotted, solid, and dashed lines correspond to the top, central, and bottom fits to NSM results, see text for details.}
\label{smcorrelation}
\end{figure}

The DGT NMEs show a very good correlation with $0\nu\beta\beta$ decay NMEs~\cite{Shimizu:2017qcy,Me_0nbb1,Me_0nbb2}, which is clearly visible in Fig.~\ref{smcorrelation} for NSM, EDF, and IBM results, while the QRPA results do not follow this correlation. In the form introduced in Ref.~\cite{Shimizu:2017qcy}, the correlation between $M^{\rm DGT}$ and $M^{0\nu\beta\beta}$ is slightly different in light and heavy nuclei, because a mass-dependent factor is introduced. The reason for the mass dependence is twofold.
First, the standard definition of the $0\nu\beta\beta$ decay NME includes a factor $R=1.2 A^{1/3}$\,fm introduced to make this NME dimensionless~\cite{En17NuclMEstatusandfuture}.
Second, the two-body NMEs of the $0\nu\beta\beta$ operator between harmonic oscillator single-particle states (the basis used by all many-body calculations presented in Fig.~\ref{smcorrelation}) is proportional to $1/b$, where $b=\sqrt{\hbar/(m_N \omega)}$ is the oscillator length. The NSM calculations use $\hbar\omega=45A^{-1/3}-25A^{-2/3}$\,MeV~\cite{hw_fit_A} from a fit to charge radii, and the same $A$ dependence is introduced in the IBM~\cite{Barea:2009zza}.
Overall, $M^{0\nu\beta\beta}$ NMEs thus include a scaling factor $R/b \sim A^{1/6}$.
Since the DGT NME is free from these dependences on the mass number, the best correlation is expected between $M^{\rm DGT}$ and $M^{0\nu\beta\beta}/A^{1/6}$.

We have tested several $A$ dependences when fitting linearly the relation between the NSM DGT and $0\nu\beta\beta$ NMEs.
In line with the above expectation, the best fit is obtained for $M^{\rm DGT}$ vs~$M^{0\nu\beta\beta}/A^{1/6}$, with the highest correlation coefficient
\begin{equation}
r_{\mathrm{NSM}}=\frac{\sum_{i}(x_i-\mu_x)(y_i-\mu_y)}{\sqrt{(\sum_{i}(x_i-\mu_x)^2)(\sum_{i}(y_i-\mu_y)^2)}}=0.90 \,.
\end{equation}
Figure~\ref{smcorrelation} shows the best fit restricted to NSM data. Extending this to include EDF and IBM results increases the correlation coefficient to $r=0.95$, confirming that the linear correlation is common to these three many-body schemes, while $M^{\rm DGT}$ vs~$M^{0\nu\beta\beta}$ or $1.2 A^{1/3}M^{\rm DGT}$ vs~$M^{0\nu\beta\beta}$ give somewhat weaker correlations. 

Next we use the correlation in Fig.~\ref{smcorrelation} to obtain EFT $0\nu\beta\beta$ decay NMEs from the EFT DGT NMEs discussed in Sec.~\ref{DGT_EFT}.
Since the latter are fitted to experimental data from GT decays or GT strengths from charge-exchange reactions, for consistency the NSM DGT NMEs need to include the usual phenomenological quenching factor $q$ to reproduce GT decay and GT strength data.
Here, we consider conservative ranges  $0.7 \leqslant q \leqslant 0.8$ for $^{48}$Ca~\cite{MartinezPinedo:1996vz} and $0.42 \leqslant q \leqslant 0.65$ for heavier nuclei~\cite{Perez:2018cly,Caurier:2011gi,KamLAND-Zen:2019imh}. Note that the NSM with quenching values in these ranges also reproduces well the measured two-neutrino $\beta\beta$ decays.

The EFT $0\nu\beta\beta$ decay NMEs are therefore obtained from the EFT DGT NMEs by
\begin{equation}
M^{0\nu\beta\beta} = A^{1/6} \, \frac{M^{\rm DGT} /q^2-n}{m}\,,
\label{correlation}
\end{equation}
where $n$ and $m$ are the parameters obtained in the linear fit.
The bottom, central, and top (dashed, solid, and dotted) lines in Fig.~\ref{smcorrelation} correspond to $n=-0.180, m=0.447$, $n=-0.106, m=0.536$, and $n=-0.056, m=0.699$, respectively, where the values are obtained as follows. For the central line, $m$ and $n$ give the best fit to the NSM results. For the slope $m$ of the top (bottom) line the same function is fitted to the ten NSM points above (below) the central line with the largest ratio of distance to the central line (which effectively weights larger NMEs somewhat more) over $M^{0\nu\beta\beta}/A^{1/6}$. The top (bottom) intercepts $n$ are determined by applying the previously obtained slopes to the same ten extreme NSM results, and choosing the $n$ value for which all NSM points lie below (above) the top (bottom) line. By construction, every NSM result lies within the top-bottom band given by the gray area in Fig.~\ref{smcorrelation}. While the NSM results cover nuclei from $A=44$ to $A=136$, the same correlation fit$_\mathrm{NSM}$ also encompasses the EDF and IBM results, extending the combined range to $44 \leqslant A \leqslant 238$ and thus for all nuclei considered in this work.

\subsection{$0\nu\beta\beta$ decay nuclear matrix elements}

\begin{figure}[t]
\centering
\includegraphics[width=\columnwidth]{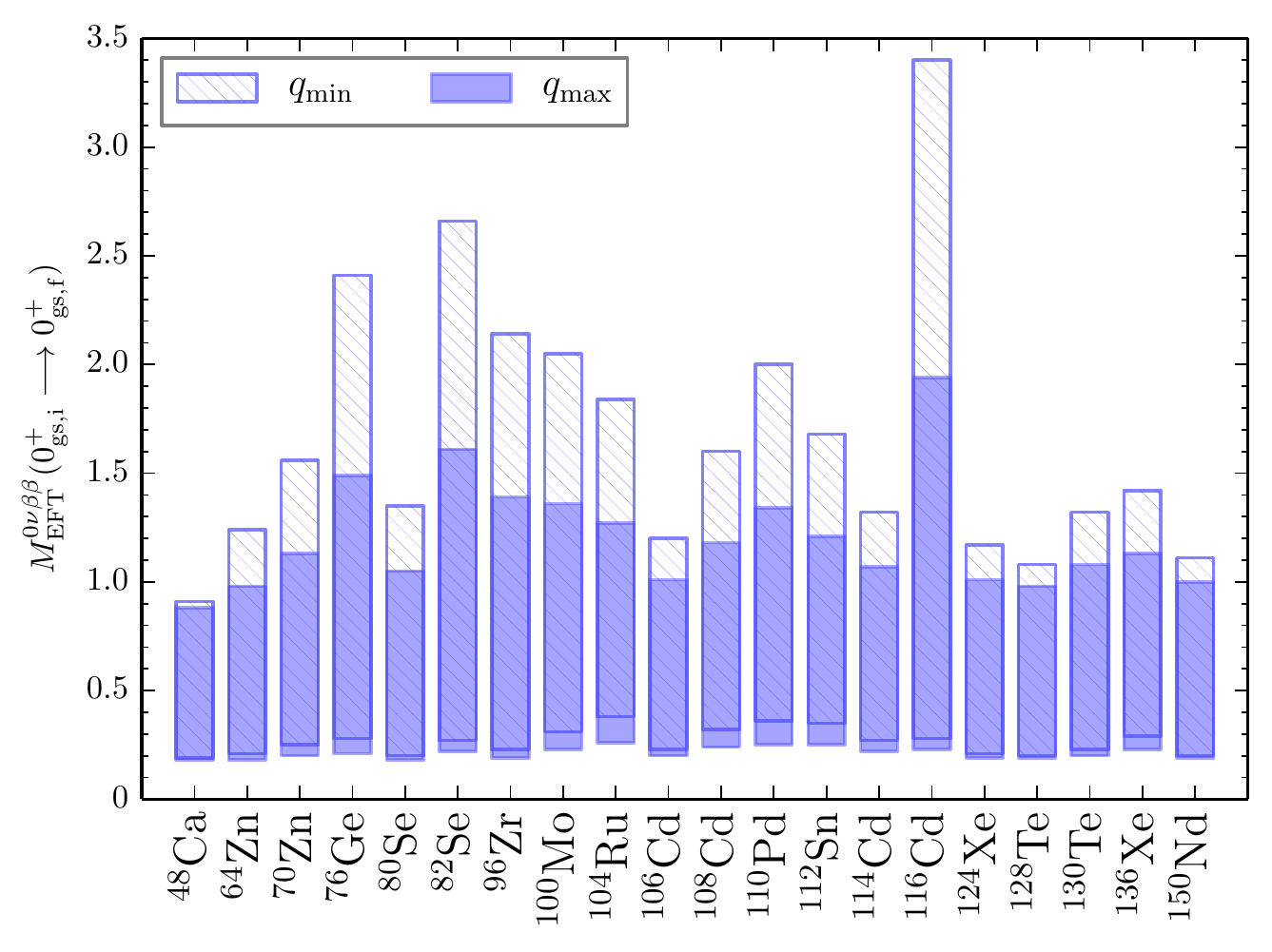}
\caption{Nuclear matrix elements for $0\nu\beta\beta$ decay predicted in the EFT for a broad range of nuclei. The hatched (filled) blue bars show predictions obtained with a quenching factor $q_\mathrm{min}=0.42$ ($q_\mathrm{max}=0.65$), except for $^{48}$Ca where $q_\mathrm{min}=0.7$ ($q_\mathrm{max}=0.8$) is used; see text for details.}
\label{fig:resultsEFT}
\end{figure}

\begin{table*}[t]
\caption{Nuclear matrix elements for $0\nu\beta\beta$ decays. The EFT DGT NMEs (column four) are obtained with the LECs ${C}_{\beta_1}$ and ${C}_{\beta_2}$ (column two and three) fitted to experimental $\log(ft)$ values or GT strengths from charge-exchange reactions for the intermediate nuclei~\cite{NDS64, NDS70, NDS80, NDS100, NDS104, NDS106, NDSlist, NDS110, NDS112, NDS114, NDS116, NDS128, BGT48, BGT76Ge, BGT82Se, BGT96, BGT130, BGT136, BGT150}, except for $^{124}\text{Xe}$ where data systematics on neighboring nuclei are used following Ref.~\cite{Perez:2018cly}. In some cases only one $\log(ft)$ value or GT strength of the investigated transition was available to obtain the LEC (e.g., for $^{76}{\rm Ge}$). The EFT $0\nu\beta\beta$ NMEs (column five and six) are based on the correlation in Eq.~(\ref{correlation}) including the uncertainties from the correlation in Fig.~\ref{smcorrelation}, the EFT truncation uncertainties in Eq.~(\ref{delta}), and considering all possible orbital combinations in Table~\ref{tab:spinparity}. The NSM quenching factors $q_\mathrm{max}=0.65$, $q_\mathrm{min}=0.42$ are used, except for $^{48}$Ca where $q_\mathrm{max}=0.8$, $q_\mathrm{min}=0.7$ are used. For comparison, we also give the range of NME results from NSM  (column seven) \cite{Mene09nbb,NSM2,NSM4,NSM3}, EDF (column eight) \cite{EDF1,EDF2}, QRPA (column nine) \cite{QRPA1,QRPA4,QRPA3,QRPA2,QRPA5}, IBM (column ten) \cite{Barea:2015kwa,IBM2} and \textit{ab initio} calculations (column eleven) (multi-reference~\cite{Yao:2019rck} and valence-space~\cite{Belley:2020ejd} IMSRG as well as coupled-cluster theory~\cite{Novario:2020dmr}).\\}
\centering
\begin{tabular}{p{0.05\textwidth}>{\centering}p{0.033\textwidth}p{0.051\textwidth}|>{\raggedleft}p{0.043\textwidth}p{0.048\textwidth}p{0.085\textwidth}|p{0.075\textwidth}p{0.09\textwidth}>{\centering}p{0.08\textwidth}>{\centering}p{0.08\textwidth}>{\centering}p{0.08\textwidth}>{\centering}p{0.08\textwidth}>{\centering\arraybackslash}p{0.08\textwidth}}
\hline\hline&&&&&&&&&\\[-3mm]
\multicolumn{3}{c|}{\multirow{2}{*}{Decay}} &
\multicolumn{1}{c}{\multirow{2}{*}{${C}_{\beta_1}$}}&
\multicolumn{1}{c}{\multirow{2}{*}{${C}_{\beta_2}$}}&
\multicolumn{1}{c|}{\multirow{2}{*}{$M^{\rm DGT}_{\rm EFT}$}} &
\multicolumn{2}{c}{$M^{0\nu\beta\beta}_{\rm EFT}$} &
\multicolumn{5}{c}{min($M^{0\nu\beta\beta}$)/max($M^{0\nu\beta\beta}$)}
\\
&&&&&&\multicolumn{1}{c}{$q_\mathrm{max}$}&\multicolumn{1}{c}{$q_\mathrm{min}$}&
\multicolumn{1}{c}{\multirow{1}{*}{NSM}} &
\multicolumn{1}{c}{EDF} &
\multicolumn{1}{c}{QRPA} &
\multicolumn{1}{c}{IBM}& \multicolumn{1}{c}{\textit{ab initio}} \\[0.5mm]
\hline\hline&&&&&&&&&\\[-3mm]
%%%%%%%%%%%%%%%48Ca
 $^{~48}{\rm Ca}$&$\longrightarrow$&$^{~48}{\rm Ti}$ &$0.603$& $0.125$ & $0.011(5)$ & $0.44({}^{+44}_{-26})$& $0.46({}^{+45}_{-27})$ &$0.30/1.12$ &$2.37/2.71$ &$0.54/0.66$ &$2.09$&0.25/0.75\\[1.5mm]
%%%%%%%%%%%%%%64Zn
 $^{~64}{\rm Zn}$&$\longrightarrow$&$^{~64}{\rm Ni}$ &$0.138$& $0.202$ & $0.009({}^{+8}_{-5})$ & $0.47({}^{+51}_{-29})$& $0.58({}^{+66}_{-37})$ & & & &&\\[1.5mm]
%%%%%%%%%%%%%%70Zn
 $^{~70}{\rm Zn}$&$ \longrightarrow$&$^{~70}{\rm Ge}$ &$0.269$& $0.176$ & $0.015({}^{+14}_{-10})$ & $0.54({}^{+59}_{-34})$& $0.73({}^{+83}_{-48})$ & &$4.60$ && &\\[1.5mm]
%%%%%%%%%%%%%%76Ge
 $^{~76}{\rm Ge}$&$\longrightarrow$&$^{~76}{\rm Se}$ &\multicolumn{2}{c}{$0.265$} &  $0.025({}^{+36}_{-18})$ & $0.63({}^{+86}_{-42})$ & $0.95({}^{+146}_{-67})$ & $2.30/3.37$ &   &  $3.12/5.57$&  5.14/6.34 & 2.05/2.23\\[1.5mm]
%%%%%%%%%%%%%%80Se
 $^{~80}{\rm Se}$&$\longrightarrow$&$^{~80}{\rm Kr}$ &$0.285$ &$0.112$ &  $0.008({}^{+11}_{-7})$ & $0.49({}^{+56}_{-31})$&$0.60({}^{+75}_{-40})$& & & &&\\[1.5mm]
%%%%%%%%%%%%%%%82Se
 $^{~82}{\rm Se}$&$\longrightarrow$&$^{~82}{\rm Kr}$ &\multicolumn{2}{c}{$0.336$} &  $0.030({}^{+39}_{-23})$ & $0.69({}^{+92}_{-47})$ &$1.06({}^{+160}_{-79})$ & $2.18/3.19$ &  $4.22/5.30$ &  $2.86/5.02$ & 4.19/5.21& 1.19/1.29 \\[1.5mm]
%%%%%%%%%%%%%%%96Zr
 $^{~96}{\rm Zr}$&$\longrightarrow$&$^{~96}{\rm Mo}$ &\multicolumn{2}{c}{$0.232$}&  $0.013({}^{+33}_{-10})$ & $0.55({}^{+84}_{-36})$&$0.73({}^{+141}_{-50})$ & &$5.65/6.37$ &$2.72/3.39$  &2.60/3.92&\\[1.5mm]
%%%%%%%%%%%%%%%100Mo
 $^{100}{\rm Mo}$&$\longrightarrow$&$^{100}{\rm Ru}$ & $0.390$&$0.313$ &  $0.021({}^{+22}_{-13})$ & $0.63({}^{+73}_{-40})$&$0.91({}^{+114}_{-60})$ & &$5.08/6.48$ &$2.44/3.90$  &3.84/5.08&\\[1.5mm]
%%%%%%%%%%%%%%%104Ru
 $^{104}{\rm Ru}$&$\longrightarrow$&$^{104}{\rm Pd}$ & $0.427$&$0.328$ &  $0.023(12)$ & $0.65({}^{+62}_{-39})$&$0.97({}^{+87}_{-59})$ & & & $4.96$&& \\[1.5mm]
%%%%%%%%%%%%%%%106Cd
 $^{106}{\rm Cd}$&$\longrightarrow$&$^{106}{\rm Pd}$ & \multicolumn{2}{c}{$0.214$} &  $0.007({}^{+5}_{-4})$ & $0.50({}^{+51}_{-30})$ &$0.59({}^{+61}_{-36})$ & & & && \\[1.5mm]
%%%%%%%%%%%%%%%108Cd
 $^{108}{\rm Cd}$&$\longrightarrow$&$^{108}{\rm Pd}$ & $0.378$&$0.276$ &  $0.017(9)$ & $0.60({}^{+58}_{-36})$ &$0.83({}^{+77}_{-51})$ & & & && \\[1.5mm]
%%%%%%%%%%%%%%%110Pd
 $^{110}{\rm Pd}$&$\longrightarrow$&$^{110}{\rm Cd}$ & $0.557$&$0.289$ &  $0.024({}^{+17}_{-13})$ & $0.66({}^{+68}_{-41})$ &$0.98({}^{+102}_{-62})$ & & & $5.69/6.52$ & 3.63/4.06&\\[1.5mm]
%%%%%%%%%%%%%%%112Sn
 $^{112}{\rm Sn}$&$\longrightarrow$&$^{112}{\rm Cd}$ & $0.496$&$0.295$ &  $0.019(9)$ & $0.62({}^{+59}_{-37})$&$0.88({}^{+80}_{-53})$ & & & & & \\[1.5mm]
%%%%%%%%%%%%%%%114Cd
 $^{114}{\rm Cd}$&$\longrightarrow$&$^{114}{\rm Sn}$ & $0.222$&$0.359$ &  $0.010(5)$ & $0.54({}^{+53}_{-32})$  &$0.68({}^{+64}_{-41})$ & & & & & \\[1.5mm]
%%%%%%%%%%%%%%%116Cd
 $^{116}{\rm Cd}$&$\longrightarrow$&$^{116}{\rm Sn}$ & $0.359$&$0.288$ &  $0.025({}^{+64}_{-19})$& $0.69({}^{+125}_{-46})$ &$1.03({}^{+237}_{-75})$  & &  $4.72/5.43$ &  $3.77/4.34$ &  2.82/2.98 & \\[1.5mm]
%%%%%%%%%%%%%%%124Xe
 $^{124}{\rm Xe}$&$\longrightarrow$&$^{124}{\rm Te}$ & $0.138$&$0.195$ &  $0.005({}^{+5}_{-3})$ &$0.49({}^{+52}_{-30})$& $0.55({}^{+62}_{-34})$  & & & &  &\\[1.5mm]
%%%%%%%%%%%%%%%128Te
 $^{128}{\rm Te}$&$\longrightarrow$&$^{128}{\rm Xe}$ & $0.184$&$0.057$ &  $0.003({}^{+3}_{-2})$ &$0.48({}^{+50}_{-29})$& $0.52({}^{+56}_{-32})$  & &4.11 &$4.56/5.08$ &4.40/4.54&\\[1.5mm]
%%%%%%%%%%%%%%%130Te
 $^{130}{\rm Te}$&$\longrightarrow$&$^{130}{\rm Xe}$ & \multicolumn{2}{c}{$0.155$}&  $0.007({}^{+7}_{-5})$ &$0.52({}^{+56}_{-32})$ & $0.62({}^{+70}_{-39})$  &$1.79/3.16$&$4.89/5.13$ & $1.37/4.37$ &3.96/4.15&\\[1.5mm]
%%%%%%%%%%%%%%%136Xe
 $^{136}{\rm Xe}$&$\longrightarrow$&$^{136}{\rm Ba}$ & \multicolumn{2}{c}{$0.223$} &  $0.012(6)$ &$0.57({}^{+56}_{-34})$ & $0.73({}^{+69}_{-44})$ & $1.63/2.39$& $4.20/4.24$& $1.11/2.91$ & 3.25/3.40&\\[1.5mm]
%%%%%%%%%%%%%%%150Nd
 $^{150}{\rm Nd}$&$\longrightarrow$&$^{150}{\rm Sm}$ & $0.208$&$0.084$ &  $0.003({}^{+3}_{-2})$ &$0.48({}^{+52}_{-29})$& $0.52({}^{+59}_{-32})$  & &$1.71/5.46$ & $2.71/3.01$  &2.47/3.57&\\[1.5mm]\hline\hline
\end{tabular}
\label{tab:predictions}
\end{table*}

\begin{figure*}[t]
\centering
\includegraphics[width=\textwidth]{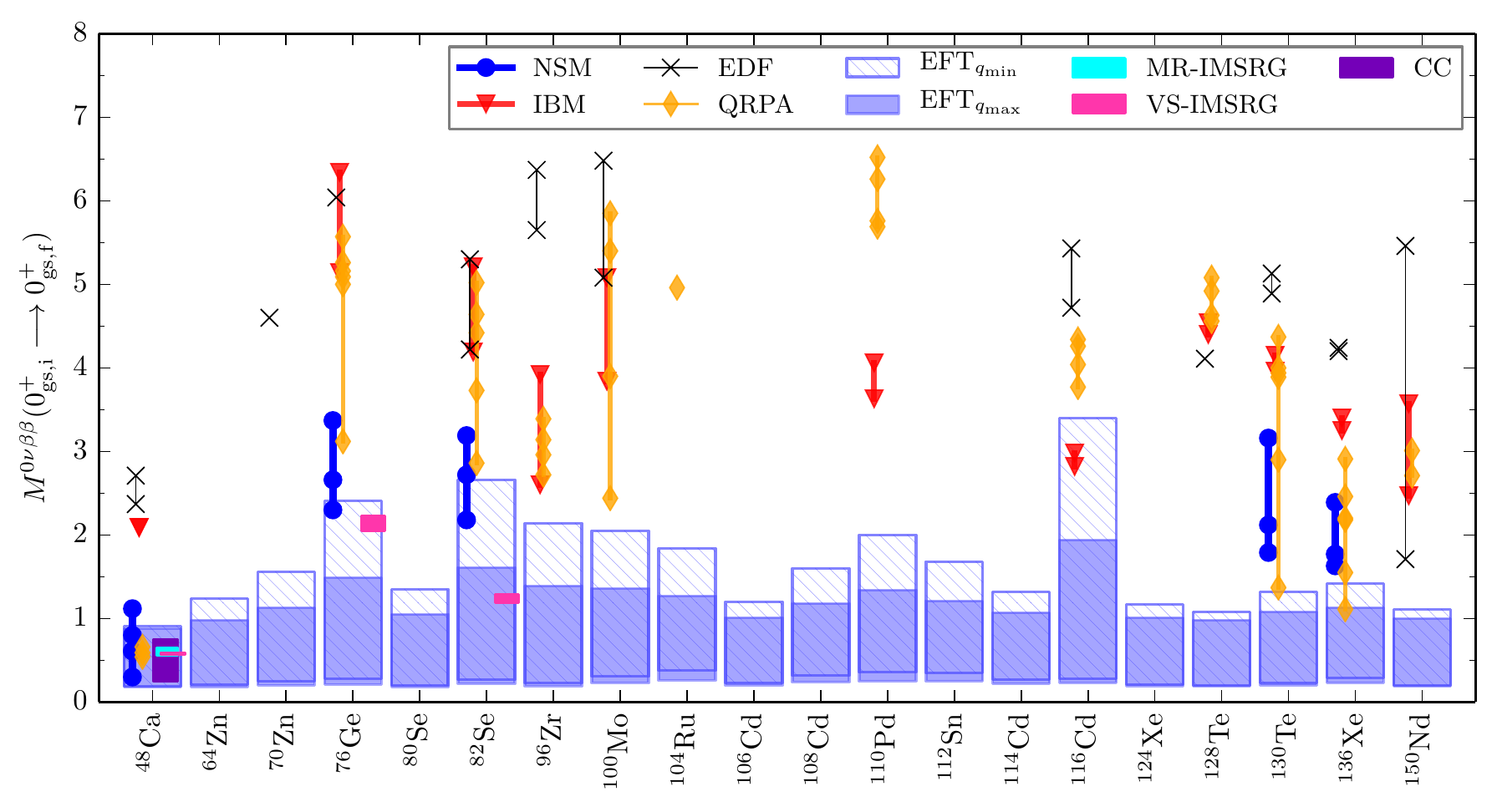}
\caption{Comparison of the EFT NMEs for $0\nu\beta\beta$ decay (same blue bars as in Fig.~\ref{fig:resultsEFT}) to results from different models, NSM (blue circles), IBM (red triangles), EDF (black crosses), and QRPA (orange diamonds), as well as recent \textit{ab initio} calculations using the MR-IMSRG (cyan bar), VS-IMSRG (pink bars), and CC theory (purple bar). For references and details see also Table~\ref{tab:predictions}. The lines between the same symbols illustrate the range of predictions within the same model.}
\label{fig:resultsall}
\end{figure*}

Table~\ref{tab:predictions} presents the EFT NMEs for all $0\nu\beta\beta$ decay candidates considered in this work, and the predicted NME ranges are shown in Fig.~\ref{fig:resultsEFT}.
For each value of the NSM quenching factor in the correlation, the central EFT $0\nu\beta\beta$ NMEs use the central fit in Fig.~\ref{smcorrelation} together with the average EFT DGT NMEs when several orbital combinations are possible. Thus for a fixed $q$ the EFT $M^{0\nu\beta\beta}$ NME range takes into account different uncertainties: (\textit{i}) the EFT truncation uncertainty of the DGT operator, Eq.~(\ref{delta}), (\textit{ii}) the possible orbital combinations entering in the DGT NME, Eq.~(\ref{eq:0to0nless}), and (\textit{iii}) the uncertainty of the NSM correlation given by the width of the gray band in Fig.~\ref{smcorrelation}. Therefore, in order to obtain the full NME uncertainty for a given $q$, we apply the DGT NMEs with EFT truncation and orbital uncertainty, see Fig.~\ref{figunc}, to the top and bottom fit lines in Fig.~\ref{smcorrelation}.

For nuclei with only one possible orbital combination, $^{48}$Ca, $^{104}$Ru, $^{108,114}$Cd, $^{112}$Sn, and $^{136}$Xe, the main source of uncertainty is associated with the width of the correlation. In these cases, the EFT truncation uncertainty has a rather small effect, as neglecting the error in the correlation reduces the corresponding uncertainty ranges to only about 15\% of the ones given in Table~\ref{tab:predictions}. For these nuclei, the uncertainty due to the NSM quenching values is also relatively small; see Fig.~\ref{fig:resultsEFT}. Likewise, for nuclei where several orbital combinations are possible but lead to a rather narrow range in or small EFT $M^\text{DGT}$ values, $^{64}$Zn, $^{106}$Cd, $^{124}$Xe, $^{128,130}$Te, and $^{150}$Nd, the orbital combinations, EFT truncation, and quenching uncertainties are also smaller compared to the dominant correlation width.

In contrast, in nuclei where several orbital combinations are possible and which exhibit a large uncertainty in the EFT $M^\text{DGT}$ predictions, $^{76}$Ge, $^{82}$Se, $^{96}$Zr, and $^{116}$Cd, this source of uncertainty is dominant over the correlation and also over the quenching uncertainty. Figure~\ref{fig:resultsEFT} illustrates that these are the nuclei with largest overall EFT $M^{0\nu\beta\beta}$ uncertainty. The overall uncertainty is also quite significant for the cases $^{70}$Zn, $^{80}$Se, $^{100}$Mo, and $^{110}$Pd, where the uncertainties due to the NSM correlation and the possible orbital combinations are comparable.

Figure~\ref{fig:resultsall} compares the EFT $0\nu\beta\beta$ NMEs with results from different model calculations using the NSM~\cite{Mene09nbb,NSM2,NSM4,NSM3}, EDF~\cite{EDF1,EDF2}, QRPA~\cite{QRPA1,QRPA4,QRPA3,QRPA2,QRPA5}, and IBM~\cite{Barea:2015kwa,IBM2}. The EFT NMEs are in general smaller than the results from phenomenological calculations. Although we make very conservative nuclear structure assumptions, we find a range $0.18 \leqslant M^{0\nu\beta\beta}_\text{EFT} \leqslant 3.40$, while phenomenological NMEs can be as large as $M^{0\nu\beta\beta}=6.5$. However, the EFT uncertainty band overlaps or is very close to the smaller predictions from the NSM and the QRPA in the case of $^{48}$Ca, as well as for the nuclei currently used in the most advanced $0\nu\beta\beta$ experiments: $^{76}$Ge, $^{82}$Se, $^{130}$Te, and $^{136}$Xe. Moreover, for $^{116}$Cd the EFT uncertainty band is consistent with the IBM result. Overall, the absolute EFT uncertainty is larger than the individual uncertainties of the other phenomenological models which often just encompass different results obtained with the same method but using different parameters. Nonetheless, in many nuclei the EFT uncertainty is smaller than the QRPA range that covers results obtained with the spherical and deformed QRPA.

We note that, although we use the same correlation found in NSM, EDF, and IBM calculations, unlike the EFT none of these approaches uses data on $\beta$ decay or GT transitions as input for the matrix element calculations. In fact, shell-model Hamiltonians are typically fine-tuned using excitation energies~\cite{RevModPhys.77.427}, energy-density functionals are determined by a fit to nuclear masses, radii and other structure information  \cite{RevModPhys.75.121, VRETENAR2005101}, and energies, radii, and electromagnetic transitions are used to fit the IBM parameters \cite{annurev.ns.31.120181.000451}, where the transition operator is mapped from fermions to bosons but without using any GT data either \cite{Barea:2009zza}. Therefore, in principle the EFT can naturally lead to different neutrinoless double-$\beta$ decay NMEs even using the same correlation with DGT transitions common to other methods.

Finally, Fig.~\ref{fig:resultsall} also shows NMEs obtained with the \textit{ab initio} multi-reference (MR) and valence-space (VS) in-medium similarity renormalization group (IMSRG)~\cite{Yao:2019rck,Belley:2020ejd} as well as coupled-cluster (CC) theory~\cite{Novario:2020dmr}. These NMEs only cover $^{48}$Ca, $^{76}$Ge, and $^{82}$Se and also prefer smaller values compared to phenomenological models. \textit{Ab initio} calculations are much more sophisticated than the EFT, but still the resulting NMEs with uncertainties completely lie within our EFT predictions, which further supports the validity of our approach. Our predictions are especially interesting for heavier nuclei beyond the reach of \textit{ab initio} frameworks, for which the only NMEs available lack quantified uncertainties.

\section{Summary}
\label{summary}

We have studied $0\nu\beta\beta$ decay within an EFT that treats nuclei as an even-even spherical collective core coupled to additional neutrons and/or protons. All microscopic details of a decay are encoded into the LECs of the effective operators. The lack of experimental data on $0\nu\beta\beta$ decay prevents fitting the LECs for this decay and, thereby, the direct prediction of $0\nu\beta\beta$ NMEs in the EFT.

In this work, we therefore followed an alternative strategy and obtained the $0\nu\beta\beta$ NMEs using a correlation between NMEs of DGT transitions and $0\nu\beta\beta$ decays. Although we have used the correlation based on NSM results, the same correlation is also supported by EDF and IBM calculations. To this end, we first calculated the NMEs for DGT transitions in the EFT and determined the LECs using experimental $ft$ values and GT strengths from charge-exchange reactions as input. The resulting DGT NMEs include uncertainty estimates both from beyond LO in the EFT and from different possible orbital combinations for the added neutrons and protons. We then used the correlation between DGT and $0\nu\beta\beta$ NMEs to predict $0\nu\beta\beta$ NMEs in the EFT with quantified uncertainties for nuclei from $^{48}$Ca to $^{150}$Nd.

The EFT $0\nu\beta\beta$ NMEs lie in a range $0.18 \leqslant M^{0\nu\beta\beta}_{\text{EFT}} \leqslant 3.40$, where the uncertainty bands for each nucleus are due to the correlation as well as the above mentioned uncertainties in the DGT NMEs. We emphasize that these ranges have been obtained for very conservative assumptions on the nuclear structure. Due to the smaller EFT DGT NME results, the EFT $0\nu\beta\beta$ NMEs are in general smaller than the predictions by the NSM, EDF, QRPA, and IBM calculations, but the EFT NME range overlaps or is close to the NSM and QRPA results for $^{48}$Ca, $^{76}$Ge, $^{82}$Se, $^{130}$Te, and $^{136}$Xe. This includes the most advanced experiments. The smaller EFT $0\nu\beta\beta$ NMEs suggest that it is important to benchmark other calculations against GT transitions. Interestingly, our EFT results are also 
consistent within uncertainties with recent \textit{ab initio} calculations for $^{48}$Ca, $^{76}$Ge, and $^{82}$Ge, which can provide future opportunities for matching the EFT for heavy nuclei to many-body calculations based on nuclear forces.

\acknowledgments

The work of C.B.~and A.S.~was supported in part by the European Research Council (ERC) under the European Union's Horizon 2020 research and innovation programme (Grant Agreement No.~101020842), the Deutsche Forschungsgemeinschaft (DFG, German Research Foundation) -- Project-ID 279384907 -- SFB 1245, and the Max Planck Society. 
The work of J.M.~was supported by the Spanish MICINN through the ``Ram\'on y Cajal'' program with grant RYC-2017-22781, the AEI ``Unit of Excellence Mar\'ia de Maeztu 2020-2023'' award CEX2019-000918-M and the AEI grant FIS2017-87534-P.
The work of E.A.C.P. was performed under the auspices of the U.S. Department of Energy by Lawrence Livermore National Laboratory under Contract No. DE-AC52-07NA27344.

\newpage
\appendix

\section{$E_{4^+}$/$E_{2^+}$ ratio and $J^P_\mathrm{gs}$\\ of initial, intermediate, and final nuclei}
\label{ratio}

\begin{table}
\caption{Initial and final nuclei (column 1 and 5) involved in DGT transitions with their ratio between the excitation energies of the first $4^+$ and $2^+$ states (column 2 and 6, respectively). The fourth column gives the ground-state spin and parity $J^P_\mathrm{gs}$ of the intermediate odd-odd nuclei (column 3). All initial and final nuclei are even-even nuclei with $0^+$ ground-state spin and parity. Data are from Ref.~\cite{NDSlist}.}
\label{tab:E24}
    \begin{center}
        \begin{tabular}{l c  | c c | l c  }
        \hline\hline
        \multicolumn{1}{c}{Initial}& \multicolumn{1}{c|}{${E_{4+}}/{E_{2+}}$}& \multicolumn{1}{c}{Intermediate}& \multicolumn{1}{c|}{$J^P_\mathrm{gs}$}&\multicolumn{1}{c}{Final}& \multicolumn{1}{c}{${E_{4+}}/{E_{2+}}$}\\\hline\hline
             $^{~48}$Ca&$1.18$  &  $^{~48}$Sc&$6^+$ & $^{~48}$Ti&$2.33$ \\
             $^{~64}$Zn&$2.33$ & $^{~64}$Cu&$1^+$&$^{~64}$Ni&$1.94$ \\
             $^{~70}$Zn&$2.02$ & $^{~70}$Ga&$1^+$& $^{~70}$Ge&$2.07$ \\
             $^{~76}$Ge&$2.50$ & $^{~76}$As&$2^-$& $^{~76}$Se&$2.38$ \\
             $^{~80}$Se&$2.55$ & $^{~80}$Br&$1^+$& $^{~80}$Kr&$2.33$ \\
             $^{~82}$Se&$2.65$ & $^{~82}$Br&$5^-$& $^{~82}$Kr&$2.34$ \\
             $^{~96}$Zr&$1.57$ &$^{~96}$Nb&$6^+$ &$^{~96}$Mo&$2.09$ \\
             $^{100}$Mo&$2.12$ & $^{100}$Tc&$1^+$& $^{100}$Ru&$2.27$ \\
             $^{104}$Ru&$2.48$ & $^{104}$Rh&$1^+$& $^{104}$Pd&$2.38$ \\
             $^{106}$Cd&$2.36$ & $^{106}$Ag&$1^+$& $^{106}$Pd&$2.40$ \\
             $^{108}$Cd&$2.38$ & $^{108}$Ag&$1^+$& $^{108}$Pd&$2.42$ \\
             $^{110}$Pd&$2.46$ & $^{110}$Ag&$1^+$& $^{110}$Cd&$2.34$ \\
             $^{112}$Sn&$1.79$ & $^{112}$In&$1^+$& $^{112}$Cd&$2.29$ \\
             $^{114}$Cd&$2.30$ & $^{114}$In&$1^+$& $^{114}$Sn&$1.68$ \\
             $^{116}$Cd&$2.37$ & $^{116}$In&$1^+$& $^{116}$Sn&$1.85$ \\
             $^{124}$Xe&$2.48$ & $^{124}$I&$2^-$& $^{124}$Te&$2.07$ \\
             $^{128}$Te&$2.01$ & $^{128}$I&$1^+$& $^{128}$Xe&$2.33$ \\
             $^{130}$Te&$1.95$ & $^{130}$I&$5^+$& $^{130}$Xe&$2.25$ \\
             $^{136}$Xe&$1.29$ & $^{136}$Cs&$5^+$& $^{136}$Ba&$2.28$ \\
             $^{150}$Nd&$2.93$ & $^{150}$Pm&$(1^-)$& $^{150}$Sm&$2.32$ \\\hline\hline
        \end{tabular}
    \end{center}
\end{table}

Table~\ref{tab:E24} lists the spin and parity of the ground state of intermediate nuclei involved in the DGT transitions studied in this work. For the initial and final even-even nuclei it gives the ratio of the excitation energies of the first $4^+$ to $2^+$ states, $E_{4^+}/E_{2^+}$.
The EFT is expected to work well for nuclei with a $E_{4^+}/E_{2^+}$ ratio between 1.5 and 2.5, a condition fulfilled for all final nuclei. However, for the initial $^{48}$Ca, $^{80,\,82}$Se, $^{136}$Xe, and $^{150}$Nd the ratio is somewhat beyond these limits. Nevertheless these systems, except for $^{150}$Nd, are doubly magic ($^{48}$Ca), semi-magic ($^{136}$Xe), or close to a semi-magic isotope ($^{84}$Se in the case of $^{80,\,82}$Se). Therefore we expect them to be well described by the EFT with a spherical core.

For transitions with an intermediate nucleus with a $1^+$ ground state, we fit the LECs to the available $\log(ft)$ values of the decay from the intermediate nucleus to the initial and final ones. When the intermediate nucleus has a ground state different from $1^+$, we fit the LECs to GT strengths extracted from charge-exchange reactions.
The only exception is the $^{124}$Xe DGT transition. The ground state of the intermediate nucleus, $^{124}$I, is a $2^-$ state. Hence, there is no GT decay from the intermediate to the initial or final nucleus. Moreover, there are no data on GT strengths available from charge-exchange reactions, requiring another strategy to fit the LECs. Here, we follow the calculation of the $2\nu$ECEC from $^{124}$Xe~\cite{Perez:2018cly}, which was based on the range of experimental EC $\log(ft)$ values from neighboring $^{122-128}$I.

\section{Neglected orbital combinations\\ in EFT calculation}\label{neglected}

\begin{table*}
    \caption{Spin and parity of various neglected neutron and proton orbitals $j_\mathrm{n}^P$ and $j_\mathrm{p}^P$, together with the energy $E$ of the lowest state with these quantum numbers and lifetime $t_{1/2}\geqslant0.1$~ns in nuclei adjacent (adj.) to the intermediate ones of a DGT transition. The spins of the adjacent odd-mass nuclei must couple to spin-parity $J^P=1^+$. Data are from Refs.~\cite{NDS69, NDS71, NDS75,NDSA77, NDS79,NDS81, NDS83, NDS105, NDS107, NDS109, NDS111, NDS113, NDS115, NDS123, NDS125, NDS129, NDS131, NDS135, NDS137, NDS149, NDS151, NDSlist}.\\} 
    \label{tab:neglected}
 \begin{tabular}{c|c r l r l| c r l r l}\hline\hline
        Initial & $j_\mathrm{n}^P$ & $E$\,[keV] & \multicolumn{1}{c}{adj.} & $E$\,[keV] & \multicolumn{1}{c|}{adj.} & $j_\mathrm{p}^P$ & $E$\,[keV] & \multicolumn{1}{c}{adj.}& $E$\,[keV] & \multicolumn{1}{c}{adj.}\\\hline\hline&&&&&&\\[-4mm]
        %%%%%%%%%%%%%%%%%70Zn
        {$^{~70}$Zn}& {${9}/{2}^+$}& 198.35 & $^{~71}$Ge& 438.64 & $^{~69}$Zn&{${9}/{2}^+$}& {1493.70} & {$^{~71}$Ga}&&\\&&&&&&\\[-4mm]\hline&&&&&&\\[-4mm]
        %%%%%%%%%%%%%%%%%76Ge
        {$^{~76}$Ge}& {${7}/{2}^+$}& 161.92 & $^{~77}$Se& 139.69 & $^{~75}$Ge&{${9}/{2}^+$}& 475.48 & $^{~77}$As & 303.92 & $^{~75}$As\\&&&&&&\\[-4mm]\hline&&&&&&\\[-4mm]
        %%%%%%%%%%%%%%%%%80Se
        {$^{~80}$Se}&{${7}/{2}^+$}& 0.~~~{} & $^{~81}$Kr& 0.~~~{} & $^{~79}$Se&{${5}/{2}^+$}& \footnotemark[1]789.40&$^{~81}$Br&  381.50 & $^{~79}$Br \\&&&&&&\\[-4mm]\hline&&&&&&\\[-4mm]
        %%%%%%%%%%%%%%%%%82Se
        {$^{~82}$Se}&{${7}/{2}^+$}& 9.41 & $^{~83}$Kr& 103.00 & $^{~81}$Se&&&&\\&&&&&&\\[-4mm]\hline&&&&&&\\[-4mm]
        %%%%%%%%%%%%%%%%%108Cd
        {$^{108}$Cd}&{${1}/{2}^+$}& $59.60$ & $^{109}$Cd& 115.74 & $^{107}$Pd&{${3}/{2}^+$}& $724.38$ & $^{109}$Ag& \footnotemark[1]1258.89 &$^{107}$Ag \\&&&&&&\\[-4mm]\hline&&&&&&\\[-4mm]
        %%%%%%%%%%%%%%%%%110Pd
        \multirow{2}{*}{$^{110}$Pd}&{${1}/{2}^+$}& 0.~~~{} & \multirow{2}{*}{$^{111}$Cd}& 113.40 & \multirow{2}{*}{$^{109}$Pd}&\multirow{2}{*}{${3}/{2}^+$}& \multirow{2}{*}{376.71} & \multirow{2}{*}{$^{111}$Ag}& \multirow{2}{*}{724.38} & \multirow{2}{*}{$^{109}$Ag} \\
        & {$({3}/{2},{5}/{2})^+$}& 736.00 & & \footnotemark[1]382.00 & \\&&&&&&\\[-4mm]\hline&&&&&&\\[-4mm]
        %%%%%%%%%%%%%%%%%112Sn
        \multirow{2}{*}{$^{112}$Sn}& {${1}/{2}^+$}& 0.~~~{} & \multirow{2}{*}{$^{113}$Sn} & 0.~~~{} & \multirow{2}{*}{$^{111}$Cd}&{${3}/{2}^+$}& 1063.93 & \multirow{2}{*}{$^{113}$In} & \footnotemark[1]1344.74&\multirow{2}{*}{$^{111}$In}\\
        &{$({3}/{2},{5}/{2})^+$}& \footnotemark[1]1042.00 &&736.00 & &{${1}/{2}^+$}& 1029.65 &&1187.62 &\\&&&&&&\\[-4mm]\hline&&&&&&\\[-4mm]
        %%%%%%%%%%%%%%%%%114Cd
        \multirow{2}{*}{$^{114}$Cd}&{${1}/{2}^+$}& 0.~~~{} & \multirow{2}{*}{$^{115}$Sn}& 0.~~~{} & \multirow{2}{*}{$^{113}$Cd}&{${3}/{2}^+$}& 828.59 & \multirow{2}{*}{$^{115}$In}& 1063.93 & \multirow{2}{*}{$^{113}$In}\\
        &{${5}/{2}^+$}&  \footnotemark[1]1734.06& & 316.21 &&{${1}/{2}^+$}& 864.14 && 1029.65 & \\&&&&&&\\[-4mm]\hline&&&&&&\\[-4mm]
        %%%%%%%%%%%%%%%%%116Cd
        \multirow{2}{*}{$^{116}$Cd}& {${1}/{2}^+$}& 0.~~~{} & \multirow{2}{*}{$^{117}$Sn}& 0.~~~{} & \multirow{2}{*}{$^{115}$Cd}&{${3}/{2}^+$}& 659.77 & \multirow{2}{*}{$^{117}$In}& 828.59 & \multirow{2}{*}{$^{115}$In}\\
        &{${3}/{2}^+$}&158.56&& \footnotemark[1]229.10 &&{${1}/{2}^+$}& 749.49&&864.14&\\&&&&&&\\[-4mm]\hline&&&&&&\\[-4mm]
        %%%%%%%%%%%%%%%%%124Xe
        \multirow{3}{*}{$^{124}$Xe}& {${1}/{2}^+$}& 0.~~~{} & \multirow{3}{*}{$^{125}$Xe}& 0.~~~{} & \multirow{3}{*}{$^{123}$Te} &{${1}/{2}^+$}& 243.38 & \multirow{3}{*}{$^{125}$I} & 148.92 & \multirow{3}{*}{$^{123}$I}\\
        &{${9}/{2}^-$}& 252.61&& \footnotemark[1]384.35&&{${3}/{2}^+$}& 188.42& & 178.02 &\\
        &{${11}/{2}^-$}& \footnotemark[1]310.55& &247.47 &&{${11}/{2}^-$}& \footnotemark[1]1084.86& &943.44 & \\&&&&&&\\[-4mm]\hline&&&&&&\\[-4mm]
         %%%%%%%%%%%%%%%%%130Te
        \multirow{2}{*}{$^{130}$Te}&{${11}/{2}^-$}& 163.93 & \multirow{2}{*}{$^{131}$Xe}& 105.51 & \multirow{2}{*}{$^{129}$Te}&\multirow{2}{*}{${9}/{2}^-, {11}/{2}^- , {13}/{2}^-$}&\multirow{2}{*}{1797.09}&\multirow{2}{*}{$^{131}$I}&&\\
        &{${9}/{2}^-$}& 341.14 &&\footnotemark[1]464.66 &&&&&&\\&&&&&&\\[-4mm]\hline&&&&&&\\[-4mm]
        %%%%%%%%%%%%%%%%%136Xe
        {$^{136}$Xe}& {${17}/{2}^-$}& 2349.10 & $^{137}$Ba&\footnotemark[1]2168.90&$^{135}$Xe&{${19}/{2}^-$}& \footnotemark[1]3303.60&$^{137}$Cs &1632.90 & $^{135}$Cs\\ \hline\hline
    \end{tabular}
\footnotetext[1]{No data on half-life available or only upper limit above 0.1\,ns.}
\end{table*}

Table~\ref{tab:neglected} lists possible neutron and proton orbitals from the adjacent odd-mass nuclei neglected in our DGT calculations. The energy of the lowest state with spin and parity $j_\mathrm{n}^P$ and $j_\mathrm{p}^P$ is also given.
We exclude orbital combinations corresponding to states with excitation energies exceeding 700\,keV in $^{70}$Zn, $^{108,\,114,\,116}$Cd, $^{110}$Pd, $^{112}$Sn, $^{124,\,136}$Xe, and  $^{130}$Te. While this threshold is somewhat arbitrary, high-energy single-particle states in the adjacent odd-mass nucleus are unlikely to contribute to the $1^+$ ground state of the intermediate odd-odd nucleus in the DGT transition.

Moreover, since the neutron $0g_{7/2}$ orbital is not part of the NSM configuration space for $^{76}$Ge and $^{80,\,82}$Se, we exclude all these $7/2^+$ states and also the ones that can only be coupled to the neglected $7/2^+$ states.
Finally, we also neglect the listed orbitals in $^{110}$Pd and $^{124}$Xe with an energy below $700$~keV because these orbitals are not expected to be very close to the Fermi level based on the NSM configuration space for these nuclei.

\bibliographystyle{apsrev4-1}
\bibliography{strongint}

\end{document}